\documentclass[printer]{rQUF2e}
\makeatletter
\def\ps@plain{\let\@mkboth\@gobbletwo\let\@oddfoot\@evenfoot\def\@oddhead{}\let\@evenhead\@oddhead}
\makeatother

\usepackage{epstopdf}
\usepackage{subfigure}

\theoremstyle{plain}

\theoremstyle{definition}
\newtheorem{definition}{Definition}

\theoremstyle{remark}

\usepackage{natbib}

\begin{document}

\title{Intrinsic Geometry of the Stock Market from Graph Ricci Flow}

\author{Bhargavi Srinivasan \thanks{Corresponding author.
Email: bhargavi.srinivasan@cnrs.fr} \\
\affil{CNRS, Laboratoire de Physique Théorique et Modèles Statistiques, Université Paris-Saclay, 91405 Orsay, France} \received{Preprint, 19 November 2026} }

\maketitle

\begin{abstract}
 We use the discrete Ollivier-Ricci curvature with Ricci flow to examine the intrinsic geometry of financial markets through the empirical correlation graph of the NASDAQ-100 index. A key feature of our approach is that we begin from an entirely ORC-spherical initial graph --- one in which all Ollivier-Ricci curvatures are strictly positive --- in direct analogy with Hamilton's (1982) foundational setting of manifolds of positive Ricci curvature. What happens to such a graph under discrete Ricci flow is an open question: no prior algorithm was designed for or tested on graphs where all curvatures are initially positive and community structure is entirely latent. We introduce ARCUS (Adaptive Ricci Curvature Unsupervised Surgery), which answers this question empirically: the Ricci flow of an ORC-spherical correlation graph develops a discrete neckpinch singularity, the curvature distribution polarising into two sharply separated populations with a self-selecting gap ratio $G \gg 1$. The procedure is genuinely recursive: each subgraph produced by surgery undergoes its own Ricci flow, following the Thurston geometrisation programme. The method is validated against the ten-stock benchmark of Mantegna (1999) and applied to the full NASDAQ-100 correlation graph, producing an eight-cluster hierarchical decomposition consistent with known market structure.
\end{abstract}


\begin{classcode}C38,C55\end{classcode}

The study of the economy as a complex dynamical system has been the focus of a great deal of attention over the past two and a half decades in the field of econophysics (\cite{jpbBook}). Stock correlations have emerged as a crucial quantity in the study of stock markets.

The empirical correlation matrix $C_{ij}$ based on the time series of returns $r_i$ for stock $i$ is computed from the Pearson correlation coefficients by the usual definition with mean $\bar{r}_i$ and variance $\sigma_i$: $  C_{ij} = \frac{< (r_i - \bar{r}_i) (r_j - \bar{r}_j)>}{\sigma_i \sigma_j}  $

We associate a distance-based edge weight $w_{ij} = \sqrt{2(1-C_{ij})}$, mapping correlations to distances with all-positive values.

A remarkable property of stock returns is their positive correlation which leads to the emergence of a market index (\cite{Laloux_1999}). The spectral properties of the stock correlation matrix have been extensively analyzed and constitute an insightful tool in the risk management of stock portfolios.

The correlation matrix can be used to construct a network with nodes given by the stocks and edge weights $w_{ij}$, thus generating a weighted, undirected, fully connected graph. The central question is how to detect community structure and hidden hierarchies in such a network. Classical partitional methods include $k$-means (\cite{macqueen1967multivariate}, \cite{lloyd1982least}) and partitioning around medoids (PAM; \cite{kaufmanrousseuw1987}). Modern network community detection methods include the modularity-maximisation approach of Girvan and Newman (\cite{Girvan}, \cite{Newman}), the Louvain method (\cite{blondel2008louvain}), the Leiden method (\cite{traag2019leiden}), Infomap (\cite{rosvall2008infomap}), and label propagation (\cite{raghavan2007label}). A natural approach in financial networks is to reduce connectivity by retaining only the most significant links. The minimum spanning tree (MST) --- the subgraph connecting all nodes with minimum total edge weight and no cycles --- was first applied to stock correlation networks by Mantegna \citeyearpar{mantegna1999hierarchical}, with further developments in \cite{tumminello2007} and \cite{Marti_2021}. MST methods have since been applied extensively to financial networks. Pruning links to create an MST changes the topology and geometry of the network, and the edge weights of the correlation matrix are very closely spaced, posing a challenge for MST methods. MacMahon and Garlaschelli \citeyearpar{macmahon2015community} showed that applying standard community detection methods directly to correlation matrices is intrinsically biased and introduced null models based on random matrix theory to correct for this. Other techniques make assumptions about the geometry of the system or the number of clusters, such as spectral clustering (\cite{Chung}).

The Ricci flow (\cite{hamilton1982three}), used with great success to solve the Poincar\'e conjecture (\cite{Perelman2002-gs}, \cite{Perelman2003-1}, \cite{Perelman2003-2}), is an intrinsic flow of the Ricci curvature tensor which, coupled with surgery, decomposes a manifold into simpler pieces. Recent advances in graph theory have made it possible to adapt this flow to discrete weighted networks, opening the door to intrinsic geometric community detection without pruning links or imposing assumptions on cluster number.

There are two steps in adapting Ricci flow to graphs, and two parallel routes. The first step is defining a discrete analogue of Ricci curvature. Two distinct approaches have been developed: combinatorial Forman-Ricci curvature (\cite{FormanRicci}), which discretises Ricci curvature via the Bochner formula on cell complexes; and probabilistic Ollivier-Ricci curvature (ORC) (\cite{Ollivier2007}, \cite{Ollivier2009-yl}, \cite{Ollivier2010-ub}, \cite{RandomGraphsORC}, \cite{LottandVillani}, \cite{Yautohoku}, \cite{Yaueigenvalues}), which defines curvature via optimal transport of random walk measures. Samal et al.\ \citeyearpar{samal2018comparative} showed the two curvatures are highly correlated in many networks; Fesser et al.\ \citeyearpar{fesser2024afrc} demonstrated an efficient Forman-Ricci alternative for community detection. The second step is devising the discrete flow. Jost and co-workers developed the Forman-Ricci route: Sreejith, Mohanraj, Jost, Saucan and Samal \citeyearpar{sreejit2016} adapted Forman-Ricci curvature to undirected networks; Weber, Jost and Saucan \citeyearpar{weber2016formanflow} introduced Forman-Ricci flow for change detection; and Weber, Saucan and Jost \citeyearpar{Weber2017-hu} characterised complex networks using Forman-Ricci curvature and associated geometric flows. Ni et al.\ \citeyearpar{ni2019community} developed the ORC route: a discrete-time algorithm updating edge weights via ORC at each iteration, explicitly building on the Jost-Weber-Saucan foundational work. Ni et al.\ were the first to develop discrete Ricci flow specifically for community detection, introducing an explicit surgery procedure applied to real network benchmarks.

Discrete curvature has also been applied directly to financial correlation networks. Sandhu, Georgiou and Tannenbaum \citeyearpar{tennenbaum2016} showed that Ollivier-Ricci curvature serves as an indicator of market fragility and systemic risk. Samal et al.\ \citeyearpar{samal2021} applied four discrete curvature measures --- including Ollivier-Ricci and Forman-Ricci --- to S\&P 500 and Nikkei 225 correlation networks over 32 years, showing that network curvature serves as an indicator of market instability. Keller-Ressel and Nargang \citeyearpar{finhyperbolic} explored the hyperbolic geometry of financial networks. In these works the network was typically MST-filtered or sparse, so community structure was partially pre-encoded. Since the appearance of this work as a preprint, ORC methods on financial networks have been further explored by Batrancea et al.\ \citeyearpar{batrancea2026} and Yang et al.\ \citeyearpar{yang2026geomherd}.

None of these algorithms has explicitly considered the case of an all-positive curvature graph --- a graph where every edge has strictly positive discrete curvature before any flow is applied. This represents a gap in the existing algorithmic landscape. All prior discrete Ricci flow algorithms were developed and tested on networks containing edges with negative curvature before any flow (\cite{ni2019community,sia2019community}), where the geometric contrast between communities is already present in the initial graph. In an all-positive curvature graph the community structure is entirely latent: there is no pre-existing geometric contrast between communities, and the flow must generate it from scratch.

It is an open question what happens to such a graph under Ricci flow. The only known analogue is from Riemannian geometry: Hamilton \citeyearpar{hamilton1982three} showed that manifolds of strictly positive Ricci curvature evolve toward a round sphere, but with heterogeneous geometry, neckpinch singularities can develop before this convergence, decomposing the manifold into simpler geometric pieces (\cite{hamilton1995singularities,AngenentKnopf2004}). Whether the Ricci flow of the NASDAQ-100 stock correlation network --- and of stock correlation networks generally, every edge of which has strictly positive ORC before any flow --- develops a discrete neckpinch that separates communities is the central question this paper seeks to answer. To this end, and to the best of our knowledge for the first time, we have developed ARCUS (\textbf{A}daptive \textbf{R}i\textbf{c}ci \textbf{C}urvature \textbf{U}nsupervised \textbf{S}urgery), a discrete Ricci flow algorithm designed to detect community structure in all-positive ORC graphs through an identification of neckpinch singularities, \emph{following the Thurston geometrisation programme} (\cite{thurston2}).

Hamilton's Ricci flow deforms the metric according to
\begin{equation}
 \frac  {\partial g_{\mu\nu} }{ \partial t} = -2 R_{\mu\nu}
\end{equation}
where $g_{\mu\nu}$ is the metric tensor and $R_{\mu\nu}$ the Ricci curvature tensor. Ollivier \citeyearpar{Ollivier2009-yl,Ollivier2010-ub} introduced the discrete analogue via optimal transport, defining the Ollivier-Ricci curvature on graph edges as follows.

The Ollivier-Ricci curvature is based on the optimal transport of probability measures associated with a lazy random walk. The discrete Ollivier-Ricci curvature $\kappa_{ij}$ on graph edge $ij \in E$, for a graph $G=(V,E,w)$ with vertices $V$, edges $E$ and weights $w$, compares the Wasserstein distance between two nodes with the shortest distance or edge weight through 
\begin{equation}
\kappa_{ij} = 1-  \frac{W(m_i,m_j)}{d(i,j)} 
 \end{equation}
 where $d(i,j)$ is the shortest distance between nodes $i$ and $j$. The Wasserstein distance is defined in the appendix and is related to the probability measures on each site. As explained in the appendix, we use a lazy random walk probability measure, with idleness parameter $\alpha=1/2$. Therefore, we drop the $\alpha$ index, with the understanding that $\alpha=1/2$ in this work.

From this definition, we see that if nodes $i$ and $j$ belong to the same community, they tend to share neighbours, so the Wasserstein distance does not exceed $d(i,j)$ --- intra-community edges have positive ORC. If nodes $i$ and $j$ belong to different communities, they tend not to share neighbours; in mixed-curvature networks the accepted wisdom assigns negative ORC to such inter-community edges. In our ORC-spherical setting, however, all edges are initially positive --- the inter-community structure is latent and must emerge from the flow. On a flat graph (Ricci-flat), the two distances are identical and $\kappa = 0$; the regular grid is an example.

We note that $\kappa = 0$ need not identify flat geometry alone: certain product manifolds also have zero curvature in one direction. The cylinder $S^{n-1} \times \mathbb{R}$ is a canonical example --- zero curvature along its axis, positive curvature in the cross-sectional sphere. This distinction will be important when we interpret the surgery links at $\kappa \approx 0$ as cylindrical rather than flat (see also Appendix~E).

\begin{definition}[ORC-spherical graph]
We call a weighted graph $G=(V,E,w)$ \emph{ORC-spherical} if $\kappa_{ij} > 0$ for all edges $(i,j) \in E$, where $\kappa_{ij}$ denotes the Ollivier-Ricci curvature of the edge under the weight-proportional random walk measure. A graph is approximately flat on an edge if $\kappa_{ij} \approx 0$, and hyperbolic on an edge if $\kappa_{ij} < 0$.
\end{definition}

With this definition of coarse curvature, we can now build the Ricci flow, to generate a time-dependent family of weighted graphs $G(V,E,w(t))$, such that the weight $w_{ij}(t)$ on edge $ij$ changes proportionally to the Ollivier-Ricci curvature $\kappa_{ij}(t)$ at time $t$.
Ollivier \citeyearpar{Ollivier2009-yl,Ollivier2010-ub} suggested a flow with continuous time $t$,
\begin{equation}
\frac  {d}{dt}{w_{e}(t)} = -\kappa_{e}(t)w_{e}(t)
 \end{equation}
where edge $e\in E$,  with Ollivier-Ricci curvature $\kappa_{e}$ and edge weight $w_{e}(t)$.
Jost and co-workers used a discrete time algorithm with the Forman-Ricci curvature \citeyearpar{weber2016formanflow,Weber2017-hu}. Sia and co-workers \citeyearpar{sia2019community} introduced an algorithm based on surgery with curvature but links were not updated. Ni and co-workers \citeyearpar{ni2019community} suggested a discrete-time algorithm while updating the edge-weights. Each iteration of this algorithm consists in updating all edge weights of the network simultaneously. The $l^{th}$ step of this process gives
 \begin{equation}
w_{ij}^{(l+1)} = (1- \kappa_{ij}^{(l)} ) w_{ij}^{(l)}
 \end{equation}
 where $w_{ij}^{(l)}$ is the weight of edge $ij$ at the $l^{th}$ iteration and $\kappa_{ij}^{(l)}$ is the Ricci curvature of edge $ij$ at the $l^{th}$ iteration. Initially $w_{ij}^{(0)} = w_{ij}$, the initial correlation-derived weights. The GraphRicciCurvature implementation (\cite{github}) uses a normalised variant in which edge weights are rescaled at each iteration to preserve the mean, preventing collapse while maintaining the curvature dynamics.

The Ricci flow alone does not produce communities --- it must be coupled with surgery. In a mixed-curvature network, inter-community edges already carry negative curvature before any flow, and the Ricci flow amplifies this pre-existing structure: positively curved intra-community links shrink while negatively curved inter-community links stretch (\cite{Weber2017-hu}, \cite{ni2019community}). In our ORC-spherical setting, no negative curvature links exist initially: the flow must generate the curvature differentiation from scratch, driven by the edge weight heterogeneity of the correlation matrix. Surgery identifies a set of edges meeting a geometric criterion and removes them, disconnecting the graph into subgraphs. Sia et al.\ \citeyearpar{sia2019community} used edge curvature alone, without any flow, as the surgery criterion. Ni et al.\ \citeyearpar{ni2019community} developed a sweeping procedure: the flow and pruning are interleaved, with links pruned repeatedly at fixed iteration intervals until the community structure emerges across the full graph. This is a single-level procedure applied to the whole graph simultaneously --- it is not recursive, and no clean gap-based surgery criterion is identified. On a mixed-curvature graph, this sweeping approach is effective because the community structure is already partially encoded in the initial geometry: repeated pruning of growing edges gradually reveals the pre-existing contrast. On an ORC-spherical graph, where all edges are initially positively curved and geometrically indistinguishable, repeated sweeping has no pre-existing contrast to exploit. A different approach is needed.

In the continuum, Hamilton \citeyearpar{hamilton1982three} showed that Riemannian manifolds of positive Ricci curvature evolve under Ricci flow toward a round sphere. However, when the initial geometry is not perfectly symmetric --- as when edge weights are heterogeneous --- neckpinch singularities can develop before this convergence, separating the manifold into simpler pieces (\cite{hamilton1995singularities,hamilton1995,AngenentKnopf2004}). In the continuum, a neckpinch singularity has locally cylindrical geometry $S^{n-1} \times \mathbb{R}$, characterised by zero curvature in the neck direction \citeyearpar{angenent2004}. In the discrete ORC-spherical setting there is no established geometric theorem guaranteeing either the existence or the form of such singularities. We therefore adopt an empirical definition: a discrete neckpinch is a population of edges whose ORC converges to near-zero ($\kappa \approx 0$) under the flow --- the discrete analogue of cylindrical geometry --- sharply separated from the positively curved intra-community population. Unlike the setting of Ni et al.\ \citeyearpar{ni2019community}, where inter-community links are already hyperbolic before any flow is applied, we have no a priori guarantee that neckpinch singularities will appear at all, nor what geometric form they will take. Identifying whether and how they emerge --- and designing a surgery criterion that exploits them --- is the central empirical aim of this paper, addressed by ARCUS.

The NASDAQ-100 index comprises 101 stocks spanning technology, healthcare, consumer discretionary, and industrial sectors, giving a fully connected weighted correlation graph on 101 nodes with $\frac{101 \times 100}{2} = 5050$ edges. Stock prices were recovered from Yahoo! Finance \cite{yahoo_fin} using the index components from \cite{wiki_nasdaq}; we compute daily log-returns and average the Pearson correlation matrix over the period 1 January 2020 to 20 November 2024. The edge weights $w_{ij} = \sqrt{2(1-C_{ij})}$ map correlations to distances, with the average pairwise correlation of $38\%$ corresponding to an average edge weight of $1.11$. The empirical NASDAQ-100 correlation matrix breaks this degeneracy. What emerges from the Ricci flow of this broken-symmetry graph is the central question of this paper.

\begin{figure}
\begin{center}
\begin{minipage}{150mm}
\subfigure{
\resizebox*{7cm}{!}{\includegraphics[width=.85\columnwidth]{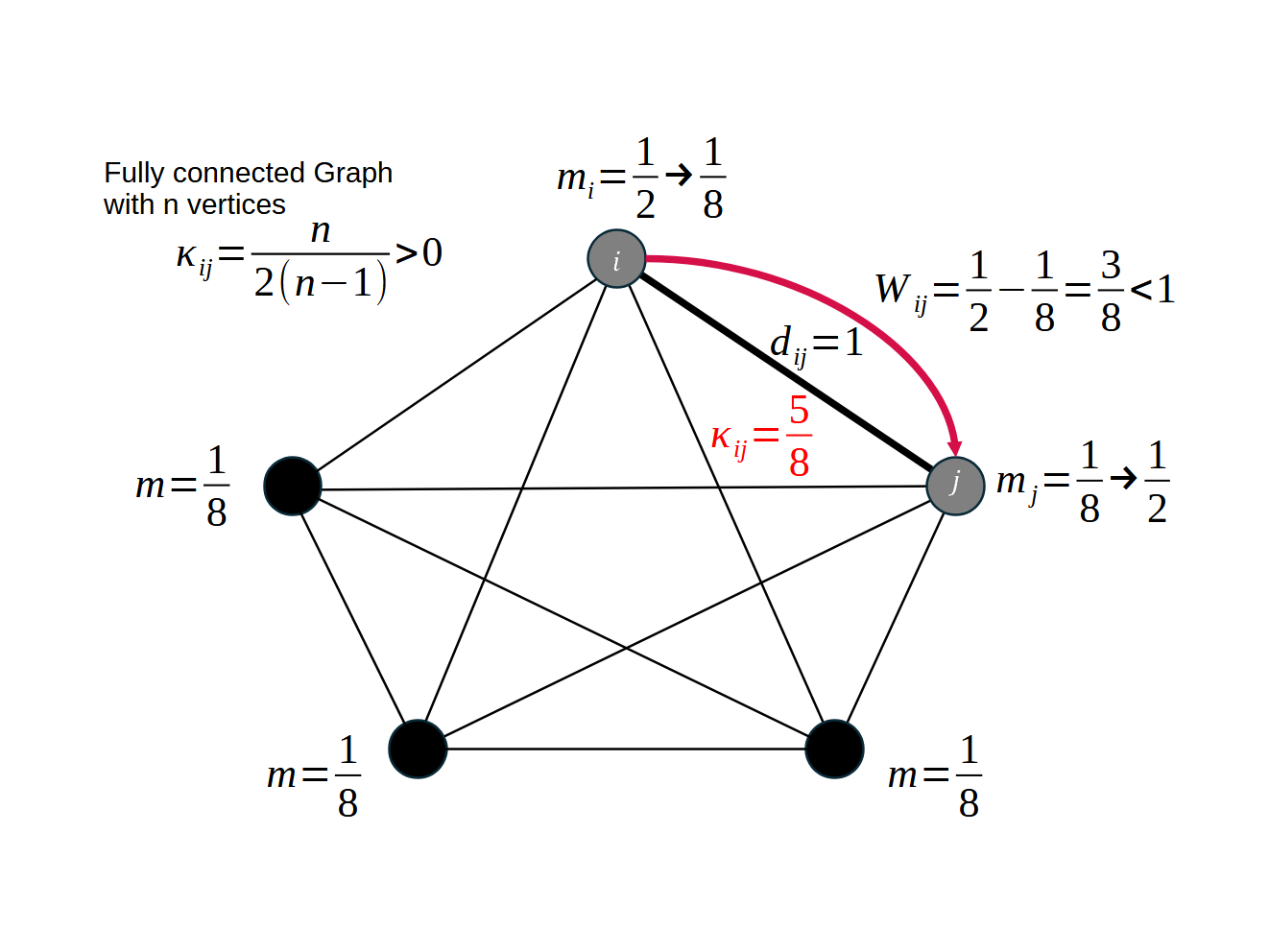}}}
\subfigure{
\resizebox*{7cm}{!}{\includegraphics[width=.85\columnwidth]{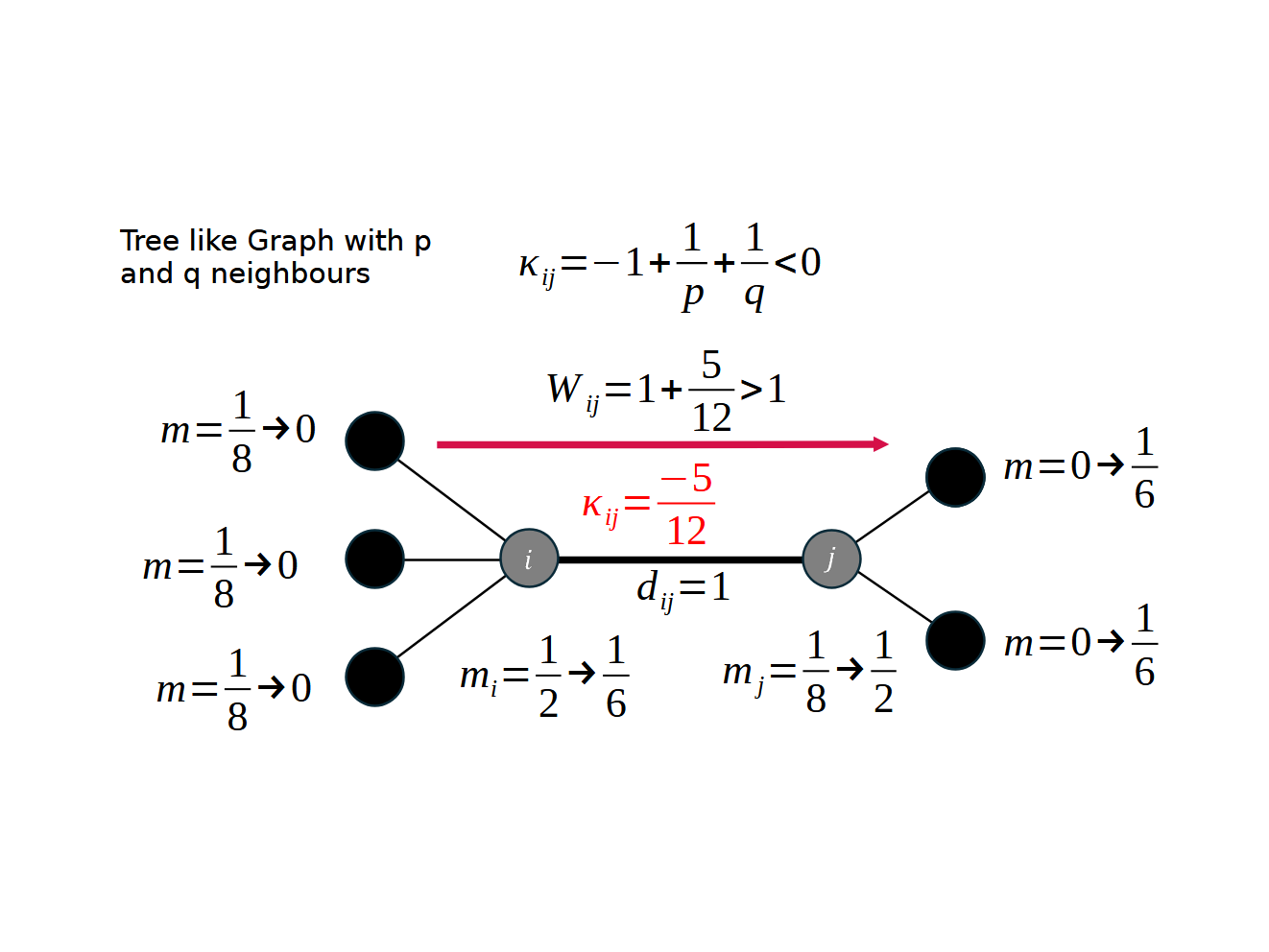}}}
\caption{\label{fig:wasssphere}Illustration of Wasserstein distance and Graph curvature on fully connected and tree-like graphs}
\end{minipage}
\end{center}
\end{figure}


A key insight into the geometry of our problem comes from first considering the fully connected graph with equal edge weights --- the degenerate limit of the empirical network. Such a graph represents a zeroth-order approximation to the correlation network, corresponding to a constant off-diagonal correlation matrix --- the discrete analogue of the market mode of Laloux et al.\ \citeyearpar{Laloux_1999}, in which a single factor drives all pairwise correlations equally. For $n=101$ nodes with $\alpha=1/2$, all Ollivier-Ricci curvatures are equal and given by $\kappa_{ij} = (1-\alpha)\frac{n}{n-1} = 0.505$: the maximally degenerate ORC-spherical case (\cite{Yautohoku}). Under Ricci flow this graph evolves self-similarly: all edges shorten at equal rates, maintaining uniform curvature throughout, and no community separation is possible.

\begin{figure}
\begin{center}
\begin{minipage}{150mm}
\subfigure{
\resizebox*{7cm}{!}{\includegraphics[width=.85\columnwidth]{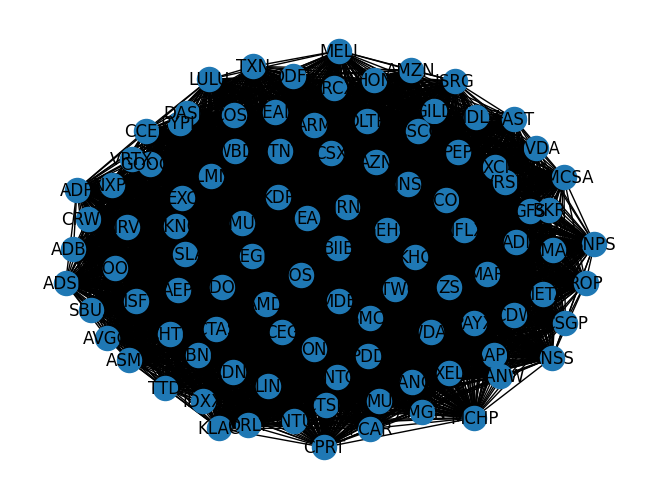}}}
\subfigure{
\resizebox*{7cm}{!}{\includegraphics[width=.85\columnwidth]{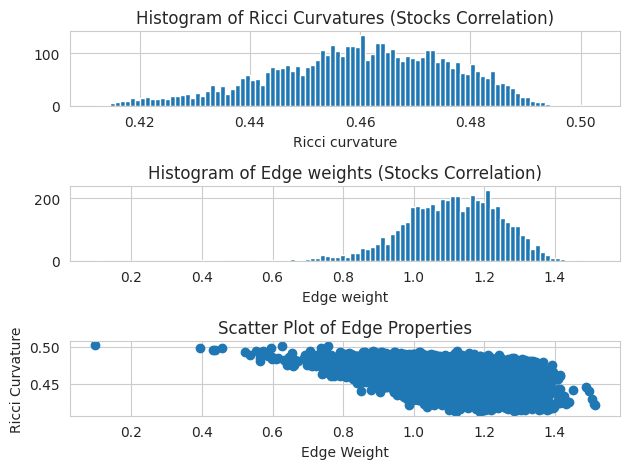}}}
\caption{\label{fig:Nasdaq100}NASDAQ-100 complete graph}
\end{minipage}
\end{center}
\end{figure}

We now turn to the full empirical network. The Ollivier-Ricci curvatures are computed using the GraphRicciCurvature library (\cite{github}) with $\alpha = 1/2$. The left panel of Fig.\ \ref{fig:Nasdaq100} shows the complete graph of 101 stocks with edges coloured by ORC value; the right panel shows the corresponding curvature histogram. The distribution is unimodal, tightly concentrated around a median of approximately $0.38$, and shifted downward from the equal-weight baseline of $0.505$ due to the heterogeneity of the empirical correlations. All 5050 edges have strictly positive ORC before any flow is applied: the NASDAQ-100 correlation graph is ORC-spherical. There is no bimodal structure, no gap, and no pre-existing separation between potential intra-community links and inter-community links --- every edge looks geometrically similar at the outset. The community structure is entirely latent.

As a natural starting point, we apply the discrete Ricci flow of Ni et al.\ \citeyearpar{ni2019community} to the NASDAQ-100 stocks correlation network. The result is ARI = 0: no community structure is recovered and no clean cut is identified. We show that ORC-spherical graphs develop a clean bimodal curvature separation under Ricci flow --- a geometric property that enables principled surgery and that previous approaches did not exploit.

Examining the scatter plot of edge curvature against edge weight after several iterations of Ricci flow, we observed an unexpected geometric structure: the curvature values began to separate into two distinct populations, while the weight distribution showed no corresponding clean separation. This was not predicted by any prior theory for ORC-spherical graphs --- it was an empirical discovery. The curvature gap, rather than the weight gap, was the signal. This geometric observation motivated the development of ARCUS: an algorithm designed to detect, quantify, and exploit the curvature gap as the surgery criterion.

ARCUS operates as follows. Given an ORC-spherical weighted graph, it runs the normalised Ricci flow, monitoring the curvature distribution at each iteration. The flow is allowed to run until the curvature distribution separates into two well-defined populations --- a positively curved intra-community population and a near-flat surgery population. Surgery is then performed along the near-zero curvature links --- the discrete neckpinch --- disconnecting the graph into subgraphs. Each subgraph undergoes its own fresh Ricci flow from its new initial curvature state and the procedure recurses. The stopping criterion is entirely data-driven: ARCUS halts when the curvature gap ratio $G$ exceeds a minimum threshold $G_{\min}$ and the surgery set is stable across three consecutive iterations, so neither the number of iterations nor the surgery threshold is chosen by the user.

The recursive structure of ARCUS is the discrete analogue of Thurston's geometrisation programme \citeyearpar{thurston2}. In Thurston's programme, successive surgeries decompose a 3-manifold until each irreducible piece carries one of eight canonical geometries. Here, successive surgeries decompose the ORC-spherical correlation graph until each piece is either too small to decompose further or produces no disconnection under Ricci flow. We observe empirically on the NASDAQ-100 that each subgraph produced by surgery has all-positive initial ORC, enabling the same geometric procedure to apply at every level; whether this is a general property of ORC-spherical graphs under surgery is an open question. No prior discrete Ricci flow community detection algorithm was genuinely recursive in this sense.

\begin{figure}
\begin{center}
\begin{minipage}{150mm}
\subfigure{
\resizebox*{4.5cm}{!}{\includegraphics[width=.3\columnwidth]{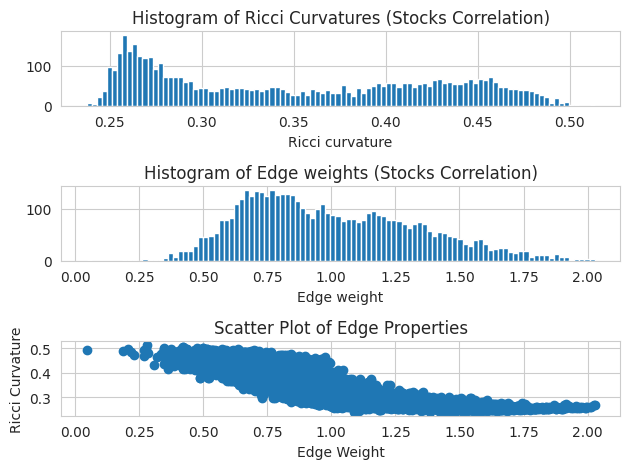}}}
\subfigure{
\resizebox*{4.5cm}{!}{\includegraphics[width=.3\columnwidth]{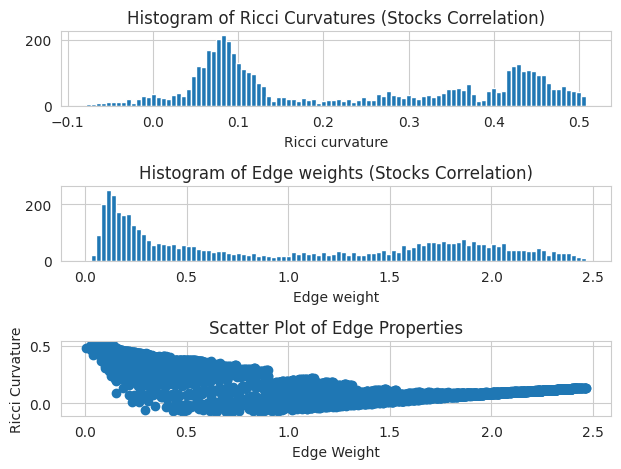}}}
\subfigure{
\resizebox*{4.5cm}{!}{\includegraphics[width=.3\columnwidth]{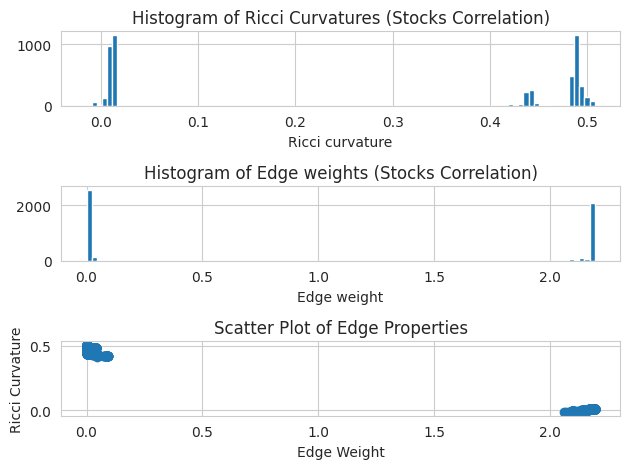}}}

\subfigure{
\resizebox*{4.5cm}{!}{\includegraphics[width=.3\columnwidth]{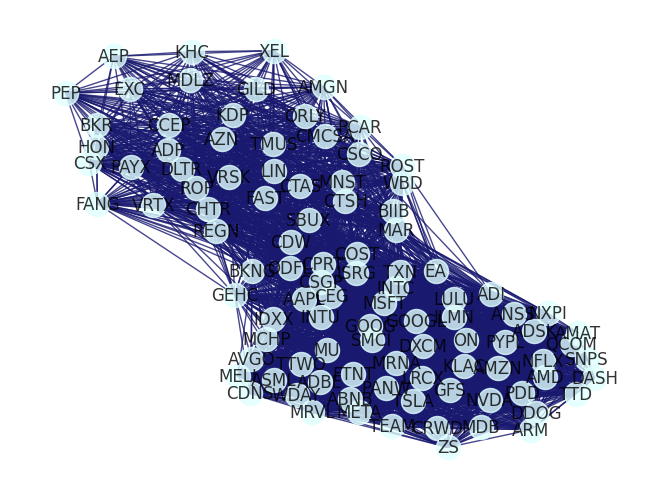}}}
\subfigure{
\resizebox*{4.5cm}{!}{\includegraphics[width=.3\columnwidth]{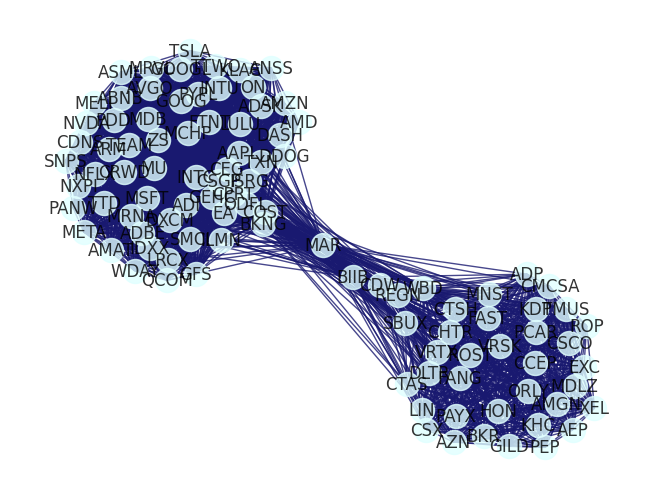}}}
\subfigure{
\resizebox*{4.5cm}{!}{\includegraphics[width=.3\columnwidth]{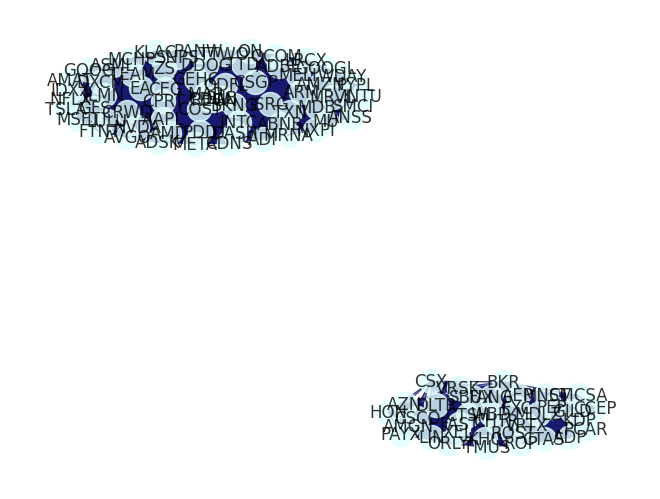}}}

\includegraphics[width=\columnwidth]{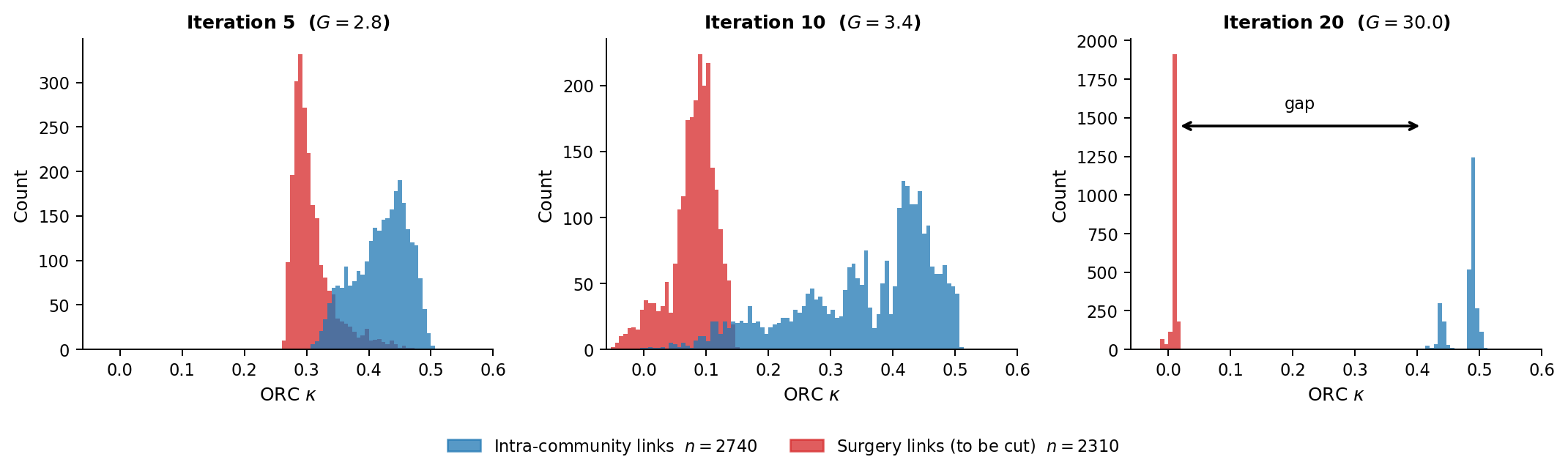}

\caption{\label{fig:Nasdaq100Level1RicciFlow}\textbf{(a)} NASDAQ-100 Ricci flow at iterations 5, 10 and 20 (left to right). Each panel shows: (top) histogram of edge curvatures $\kappa$; (middle) histogram of edge weights; (bottom) scatter plot of curvature vs.\ edge weight. \textbf{(b)} ORC distributions at the same three iterations with edges coloured by the final surgery assignment: blue = intra-community ($n=2740$, positively curved), red = surgery links ($n=2310$, to be cut). The red surgery links are the long, flat links ($\kappa \approx 0$, large edge weight) that form the discrete neckpinch --- the cylindrical geometry separating communities. At iteration 5 the two populations almost completely overlap ($G \approx 0.3$). At iteration 10 separation is developing ($G \approx 4.4$). By iteration 20 the gap is fully open and $G = 30.5 \gg 1$, confirming the surgery set is self-selecting (Appendix~C).}

\end{minipage}
\end{center}
\end{figure}

%

We quantify the sharpness of the curvature separation using the \textbf{curvature gap ratio}
\begin{equation}
G = \frac{\mu(\kappa^+) - \mu(\kappa^-)}{\sigma_{\mathrm{pooled}}}
\end{equation}
where $\mu(\kappa^+)$ is the mean ORC of the intra-community population, $\mu(\kappa^-)$ is the mean ORC of the surgery population, and $\sigma_{\mathrm{pooled}}$ is the pooled standard deviation of both populations. $G \gg 1$ indicates that the two populations are sharply separated and the surgery criterion is self-selecting: any threshold in the empty gap between the two modes yields the same surgery set. $G$ is an unsupervised analogue of the curvature gap introduced by Gosztolai and Arnaudon \citeyearpar{gosztolai2021unfolding} and used by Fesser et al.\ \citeyearpar{fesser2024afrc} as a supervised measure requiring knowledge of the ground truth partition; here the post-flow curvature population replaces the known community labels. ARCUS (\textbf{A}daptive \textbf{R}i\textbf{c}ci \textbf{C}urvature \textbf{U}nsupervised \textbf{S}urgery) brings these elements together: it is \textbf{adaptive} in that the stopping criterion and surgery threshold emerge from the data via $G$-stability rather than being fixed by the user; it operates on \textbf{Ricci curvature} via the ORC flow; it is \textbf{unsupervised} in that no labels, cluster count, or threshold parameter are required; and it performs \textbf{surgery} along the discrete neckpinch identified by the curvature gap. ARCUS runs until $G$ exceeds a minimum threshold $G_{\min}$ and the surgery set is stable across consecutive iterations (Appendix~B).

We carried out the Ricci flow on the NASDAQ-100 with $\alpha=\frac{1}{2}$ and a discrete time step (\cite{github}).\footnote{Sia et al.\ \citeyearpar{sia2019community} used edge curvature alone without Ricci flow, with the argument that triangles are a good index of community formation; we find that the full flow is necessary to open the curvature gap cleanly.} After 5 iterations the weight distribution begins to flatten and two emerging curvature peaks are visible; a scatter plot of curvature against edge weight is widely dispersed, confirming that curvature and weight are not simply proportional after the flow. After 10 iterations the dumbbell is pronounced and the two populations are beginning to separate, but $G \approx 4.4$ is not yet sufficient for surgery. By 20 iterations the curvature distribution has split into two clearly separated modes with $G = 30.5 \gg 1$, $\mu(\kappa^+) \approx 0.48$ and $\mu(\kappa^-) \approx 0.01$. Surgery is performed along the zero-curvature links --- the lower mode of the bimodal curvature distribution --- not on edge weight.

The central empirical finding of this paper is the development of a discrete neckpinch under Ricci flow. The number of iterations is not chosen by the user: ARCUS stops when $G > G_{\min}$ and the surgery set is stable across three consecutive iterations, so the convergence point emerges unsupervised from the data. At Level 1 this occurs at 20 iterations. After convergence, the ORC distribution has polarised into two sharply separated populations: a positively curved intra-community population with $\kappa \approx 0.48$ and a near-flat surgery population with $\kappa \approx 0.01$, consistent with the cylindrical geometry ($\kappa = 0$) of the continuum neckpinch. The gap between the two populations is empty, confirming the empirical definition of the discrete neckpinch proposed above. Values of $G$ range from 16 to 70 across the hierarchical levels reported in Table~\ref{tab:curvature_gap}, well above the threshold $G_{\min}=15$, confirming that the stopping criterion fires cleanly and is not sensitive to the precise choice of $G_{\min}$ within a reasonable range. We observe that the surgery population converges to $\kappa \approx 0.01$ at Level 1, with occasional values slightly below zero appearing across hierarchical levels --- within the bounds of numerical precision of the finite-precision optimal transport computation. This distinction does not affect the algorithm --- the surgery criterion depends only on the bimodal gap and is robust to small perturbations around zero. Whether the surgery links approach $\kappa = 0$ from above throughout the flow --- without ever becoming hyperbolic --- is an open question; if confirmed, it would establish a discrete analogue of Hamilton's theorem that Ricci flow on an ORC-spherical graph preserves positive curvature while developing a flat neckpinch (Appendix~E).

As the Ricci flow proceeds, the network evolves geometrically from a near-uniform sphere toward a dumbbell configuration: two dense, positively curved communities gradually become visible, connected by an increasingly flat neck of inter-community links. While the dumbbell geometry provides an intuitive visual confirmation, the identification of the neckpinch and the surgery criterion are based entirely on the curvature gap ratio $G$ --- the network layout plays no role in the algorithm. Each subgraph produced by surgery is observed to be ORC-spherical on the NASDAQ-100, and the procedure repeats.

The NASDAQ-100 is a tech-heavy index. After the first level of Ricci flow and surgery, the correlation graph splits into two communities of 66 and 35 stocks, corresponding to a technology-driven cluster and a defensive non-cyclical cluster. The geometric separation is clean: the two branches are identified entirely from curvature dynamics, without sector labels or prior classification.

Within the tech branch, Level 2 surgery identifies 16 semiconductor stocks (Table~\ref{tab:level2_semicond}) --- chipmakers, equipment manufacturers, and foundries --- recovering a well-defined industrial grouping that ARCUS finds purely through return co-movement geometry. Level 3 separates a cross-sector cluster of 10 stocks (Table~\ref{tab:level3_outlier}) with no dominant GICS theme: these are high-growth, high-variance names from consumer discretionary, IT, and communications services whose geometric cohesion reflects a latent common factor --- likely exposure to the post-COVID digital acceleration and subsequent rate-sensitivity --- rather than industrial similarity. The remaining 40-stock cluster then splits: 6 stocks including four biotechs together with Marriott and Supermicro (Table~\ref{tab:level4_biotech}), united by elevated return variance during 2020--2024 that separates them geometrically from the main growth cohort. The final split produces a 12-stock Big Tech cluster and a 22-stock growth cluster led by Tesla. The Big Tech cluster contains six of the Magnificent Seven --- Apple, Microsoft, Nvidia, Amazon, Meta, and Alphabet --- plus AMD, Adobe, Cadence, Synopsys, and Ansys. Tesla, the seventh member of the Magnificent Seven, appears in the growth cluster rather than Big Tech, reflecting its distinct return profile during the period: higher volatility, stronger co-movement with consumer discretionary and speculative growth names than with the large-cap software and platform stocks. Nvidia appears in Big Tech rather than the semiconductor cluster, consistent with its role as an AI platform company during 2020--2024 rather than a pure semiconductor play.

\begin{figure}
\begin{center}
\begin{minipage}{150mm}
\includegraphics[width=.9\columnwidth]{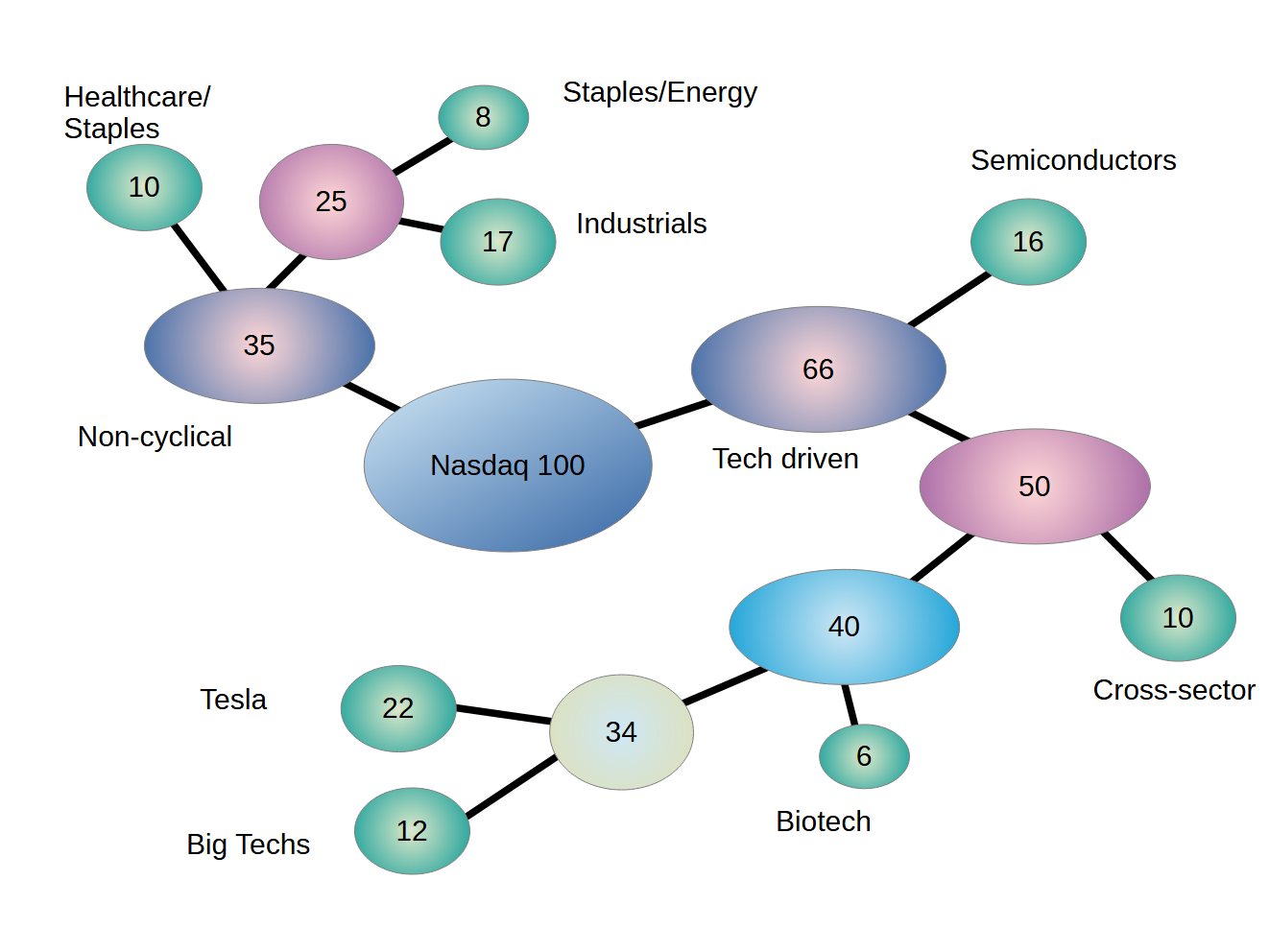}
\caption{\label{fig:Nasdaq100Level5Cluster}Full cluster hierarchy of the NASDAQ-100 produced by ARCUS (Adaptive Ricci Curvature Unsupervised Surgery, Appendix~B). At each level the flow runs until $G > G_{\min} = 15$ and the surgery set $n_{\mathrm{low}}$ is unchanged for three consecutive iterations. Every split is binary. The weight-based surgery criterion of Ni et al.\ does not produce clean splits on the dense complete correlation graph; the curvature gap criterion is used exclusively throughout the hierarchy.}\end{minipage}
\end{center}
\end{figure}

The 35-stock non-cyclical cluster represents the defensive half of the NASDAQ-100 --- lower beta, less tech-exposed names whose correlations are driven by macro factors rather than technology cycles. Level 2 surgery separates a 10-stock healthcare and utilities cluster (Table~\ref{tab:level2_healthstaples}) combining large-cap biotechs with regulated utilities and consumer staples --- a grouping united by defensive, income-oriented return characteristics. The remaining 25 stocks split into a staples/energy/media cluster of 8 (Table~\ref{tab:level3_staplesenergy}) and an industrials cluster of 17 (Table~\ref{tab:level3_industrials}). The industrials cluster is notably diverse --- spanning HR services, rail transport, industrial gases, beverages, and wireless telecoms --- but geometrically coherent: these are stable, cash-generative businesses whose return co-movements during 2020--2024 reflect exposure to the domestic economic cycle rather than technology or defensive macro factors. The full hierarchy is shown in Fig.~\ref{fig:Nasdaq100Level5Cluster} and the detailed cluster membership is listed in the appendix tables.

 The cluster assignments presented here are intended to illustrate the geometric method; detailed comparisons with MST, PCA and other methods, a broader econophysical analysis including a larger and less tech-heavy index such as the S\&P 500, and behaviour around market crashes, are the subject of a companion paper (\cite{Bhargavicluster}).
 
 In conclusion, we study the NASDAQ-100 stock index from the point of view of a fully connected network and make three contributions. ARCUS is designed for ORC-spherical weighted graphs, of which financial correlation matrices are a natural example.

Our first contribution is to study Ricci flow community detection beginning from an ORC-spherical initial graph --- the fundamental case for Ricci flow theory, in direct analogy with Hamilton's \citeyearpar{hamilton1982three} foundational setting of manifolds of strictly positive curvature. Prior work began from mixed-curvature graphs where community structure was pre-encoded; our flow must generate the geometric structure from scratch. Stock correlation networks naturally reside in this ORC-spherical regime.

Our second contribution is the identification of the neckpinch via curvature (see Appendix~E for the geometric foundations). In Ni et al.\ \citeyearpar{ni2019community} the inter-community edges already carry negative curvature before any flow; in our ORC-spherical graph the neckpinch must develop dynamically from the equal-curvature initial state, driven by edge weight heterogeneity --- a result not guaranteed a priori. A complete graph with equal edge weights undergoes Ricci flow self-similarly, collapsing to a point with no neckpinch; the development of a neckpinch before this collapse is therefore the key non-trivial finding. We show that after sufficient Ricci flow iterations the curvature distribution polarises into two populations sharply separated by an empty gap with $G \gg 1$, and the surgery set is self-selecting.

Our third contribution is the genuinely recursive surgery procedure: each community undergoes its own Ricci flow from a new initial state observed to be ORC-spherical on the NASDAQ-100, in analogy with Thurston's geometrisation programme. This yields five hierarchical levels of cluster structure in the NASDAQ-100, recovering technology, semiconductor, healthcare, staples, and energy groupings. A cross-sector cluster at level 3 suggests a latent common factor not captured by industrial classification (\cite{Bhargavicluster}).

Validation against the Mantegna \citeyearpar{mantegna1999hierarchical} ten-stock benchmark (see supplementary material) confirms that the method recovers the same sector groupings --- energy, technology, and financial --- as average-linkage hierarchical clustering in two hierarchical steps. Our curvature-based approach provides an attractive alternative to MST to study the geometric properties of the stock market, and ARCUS provides a novel, curvature-based criterion to perform surgery on highly-connected graphs with initial positive curvature. We note that the edge weight gap opens cleanly at convergence and yields identical community assignments on the NASDAQ-100; the relationship between curvature-based and weight-based surgery criteria on ORC-spherical graphs, and the extension to mixed-curvature settings where the curvature gap may not open cleanly, are directions for future work.

\section*{Acknowledgements}
We wish to acknowledge the support of the CNRS and LPTMS. Earlier versions of this work were presented at the Indo-French meet on \textit{Classical and Quantum Dynamics in Out of Equilibrium Systems} at the International Centre for Theoretical Sciences (ICTS), Bangalore (December 2024) and at the workshop \textit{The Lush World of Random Matrices} at the Zentrum f\"{u}r interdisziplin\"{a}re Forschung (ZiF), Bielefeld University, Germany (October 2025).

\section*{Disclosure of interest}
There are no relevant financial or non-financial competing interests to report.

\bibliographystyle{rQUF}
\bibliography{StocksRicciClustering}

\newpage
\section*{Tables}

\begin{table}[h!]
\scriptsize
    \centering
    \begin{tabular}{|l|l|l|l|}
    \hline
    \textbf{Ticker} & \textbf{Name}            & \textbf{Sector} & \textbf{Sub-Industry}          \\ \hline
    KLAC            & KLA Corporation           & IT             & Semiconductors Equip      \\ \hline
    ADI             & Analog Devices            & IT             & Semiconductors                \\ \hline
    LRCX            & Lam Research              & IT             & Semiconductors Equip      \\ \hline
    AMAT            & Applied Materials         & IT             & Semiconductors Equip      \\ \hline
    MCHP            & Microchip Technology      & IT             & Semiconductors                \\ \hline
    ARM             & ARM Holding               & IT             & Semiconductors                \\ \hline
    ASML            & ASML Holdings             & IT             & Semiconductors Equip      \\ \hline
    AVGO            & Broadcom                  & IT             & Semiconductors                \\ \hline
    MRVL            & Marvell Technology        & IT             & Semiconductors                \\ \hline
    MU              & Micron Technology         & IT             & Semiconductors                \\ \hline
    NXPI            & NXP Semiconductors        & IT             & Semiconductors                \\ \hline
    ON              & Onsemi                    & IT             & Semiconductors                \\ \hline
    QCOM            & Qualcomm                  & IT             & Semiconductors                \\ \hline
    GFS             & GlobalFoundries           & IT             & Semiconductors                \\ \hline
    TXN             & Texas Instrument          & IT             & Semiconductors                \\ \hline
    INTC            & Intel                     & IT             & Semiconductors                \\ \hline
    \end{tabular}
    \caption{Level2 Cluster: Semiconductor Industry}
    \label{tab:level2_semicond}
\end{table}

\begin{table}[h!]
\scriptsize
    \centering
    \begin{tabular}{|l|l|l|l|}
        \hline
        \textbf{Ticker} & \textbf{Name} & \textbf{Sector} & \textbf{Sub-Industry} \\ \hline
        ABNB & Airbnb & Cons Disc & Hotel, resort Cruise \\ \hline
        MRNA & Moderna & Health Care & Biotech \\ \hline
        PDD & PDD Holding & Cons Disc & Broadline retail \\ \hline
        TEAM & Atlassian & IT & Appli software \\ \hline
        CRWD & Crowdstrike & IT & System software \\ \hline
        TTD & Trade Desk & Comm Servic & Advert \\ \hline
        DASH & DoorDash & Cons Disc & Specialized cons services \\ \hline
        DDOG & Datadog & IT & Appli software \\ \hline
        ZS & Zscaler & IT & Appli software \\ \hline
        MDB & MongoDB & IT & System software \\ \hline
    \end{tabular}
    \caption{Level3 Cluster: Cross-sector cluster}
    \label{tab:level3_outlier}
\end{table}

\begin{table}[h!]
\scriptsize
    \centering
    \begin{tabular}{|l|l|l|l|}
    \hline
    \textbf{Ticker} & \textbf{Name} & \textbf{Sector} & \textbf{Sub-Industry} \\ \hline
    GEHC & GE HealthCare & Health Care & Health Care Equip \\ \hline
    CEG & Constellation Energy & Utilities & Electric Utilities \\ \hline
    REGN & Regeneron Pharmaceuticals & Health Care & Biotechnology \\ \hline
    BIIB & Biogen & Health Care & Biotechnology \\ \hline
    MAR & Marriott International & Cons Disc & Hotels, Resorts, Cruise Lines \\ \hline
    SMCI & Super Micro Computer & IT & Tech Hardware, Storage \& Peripherals \\ \hline
    \end{tabular}
    \caption{Level 4 Cluster: Biotech and high-variance outliers (6 stocks). The curvature gap criterion groups MAR and SMCI with the biotech stocks; both had elevated return variance during 2020--2024 that distinguishes them geometrically from the main growth cluster.}
    \label{tab:level4_biotech}
\end{table}

\begin{table}[h!]
\scriptsize
    \centering
    \begin{tabular}{|l|l|l|l|}
        \hline
        \textbf{Ticker} & \textbf{Name} & \textbf{Sector} & \textbf{Sub-Industry} \\ \hline
        AAPL & Apple Inc. & IT & Tech Hardw, Stora, Periph \\ \hline
        ADBE & Adobe Inc. & IT & Application Software \\ \hline
        AMD & Advanced Micro Dev & IT & Semiconductors \\ \hline
        AMZN & Amazon & Cons Disc & Broadline Retail \\ \hline
        ANSS & Ansys & IT & Application Software \\ \hline
        CDNS & Cadence Design Sys & IT & Application Software \\ \hline
        GOOG & Alphabet (C) & Comm Serv & Media \& Services \\ \hline
        GOOGL & Alphabet (A) & Comm Serv & Media \& Services \\ \hline
        META & Meta Platforms & Comm Serv & Media \& Services \\ \hline
        MSFT & Microsoft & IT & Systems Software \\ \hline
        NVDA & Nvidia & IT & Semiconductors \\ \hline
        SNPS & Synopsys & IT & Application Software \\ \hline
    \end{tabular}
    \caption{Level 5 Cluster: Big Tech (12 stocks). The Magnificent Seven (Apple, Microsoft, Nvidia, Amazon, Meta, Alphabet) plus their closest software and semiconductor peers. Nvidia appears here rather than in the semiconductor cluster (Level 2), reflecting its broader market role during the AI boom.}
    \label{tab:level5_bigtech}
\end{table}

\begin{table}[h!]
\scriptsize
    \centering
    \begin{tabular}{|l|l|l|l|}
        \hline
        \textbf{Ticker} & \textbf{Name} & \textbf{Sector} & \textbf{Sub-Industry} \\ \hline
TSLA & Tesla, Inc. & Cons Disc & Auto Manufacturers \\ \hline
ADSK & Autodesk & IT & Application Software \\ \hline
BKNG & Booking Holdings Inc. & Cons Disc & Hotels, Resorts, Cruise Lines \\ \hline
CDW & CDW Corporation & IT & Tech Distributors \\ \hline
COST & Costco & Cons Staples & Merch Retail \\ \hline
CPRT & Copart & Industrials & Diversified Support Serv \\ \hline
CSGP & CoStar Group & Real Estate & Real Estate Serv \\ \hline
DXCM & DexCom & Health Care & Health Care Equip \\ \hline
EA & Electronic Arts & Comm Serv & Home Entertainment \\ \hline
FTNT & Fortinet & IT & Systems Software \\ \hline
IDXX & Idexx Laboratories & Health Care & Health Care Equip \\ \hline
ILMN & Illumina, Inc. & Health Care & Life Sciences Tools \& Serv \\ \hline
INTU & Intuit & IT & Application Software \\ \hline
ISRG & Intuitive Surgical & Health Care & Health Care Equip \\ \hline
LULU & Lululemon Athletica & Cons Disc & Apparel, Accessories \\ \hline
MELI & MercadoLibre & Cons Disc & Broadline Retail \\ \hline
NFLX & Netflix & Comm Serv & Movies \& Entertainment \\ \hline
ODFL & Old Dominion Freight Line & Industrials & Cargo Ground Transp \\ \hline
PANW & Palo Alto Networks & IT & Systems Software \\ \hline
PYPL & PayPal & Financials & Transaction \& Payment Serv \\ \hline
TTWO & Take-Two Interactive & Comm Serv & Home Entertainment \\ \hline
WDAY & Workday, Inc. & IT & Application Software \\ \hline
    \end{tabular}
    \caption{Level 5 Cluster: Tesla/growth group (22 stocks). High-growth consumer discretionary and software stocks spanning multiple sectors, united by their return co-movement during the 2020--2024 period.}
    \label{tab:level5_tesla}
\end{table}

\begin{table}[h!]
\scriptsize
    \centering
    \begin{tabular}{|l|l|l|l|}
        \hline
        \textbf{Ticker} & \textbf{Name} & \textbf{Sector} & \textbf{Sub-Industry} \\ \hline
VRTX & Vertex Pharmaceuticals & Health Care & Biotechnology \\ \hline
AEP & American Electric Power & Utilities & Electric Utilities \\ \hline
AMGN & Amgen & Health Care & Biotechnology \\ \hline
AZN & AstraZeneca & Health Care & Pharmaceuticals \\ \hline
XEL & Xcel Energy & Utilities & Multi-Utilities \\ \hline
EXC & Exelon & Utilities & Electric Utilities \\ \hline
GILD & Gilead Sciences & Health Care & Biotechnology \\ \hline
KHC & Kraft Heinz & Cons Staples & Packaged Foods \& Meats \\ \hline
MDLZ & Mondelez International & Cons Staples & Packaged Foods \& Meats \\ \hline
PEP & PepsiCo & Cons Staples & Soft Drinks \& Beverages \\ \hline
    \end{tabular}
    \caption{Level2 Cluster: Health Care/Staples}
    \label{tab:level2_healthstaples}
\end{table}

\begin{table}[h!]
\scriptsize
    \centering
    \begin{tabular}{|l|l|l|l|}
        \hline
        \textbf{Ticker} & \textbf{Name} & \textbf{Sector} & \textbf{Sub-Industry} \\ \hline
    BKR & Baker Hughes & Energy & Oil-Gas Equip/Services \\ \hline
    CCEP & Coca-Cola Europacific & Cons Staples & Soft Drinks \& Beverages \\ \hline
    CHTR & Charter Comms & Comm Services & Cable \& Satellite \\ \hline
    CMCSA & Comcast & Comm Services & Cable \& Satellite \\ \hline
    DLTR & Dollar Tree & Cons Staples & Merch Retail \\ \hline
    FANG & Diamondback Energy & Energy & Oil-Gas Explo/Production \\ \hline
    ROST & Ross Stores & Cons Disc & Apparel Retail \\ \hline
    WBD & Warner Bros. Discovery & Comm Services & Broadcasting \\ \hline
    \end{tabular}
    \caption{Level3 Cluster: Staples/Energy}
    \label{tab:level3_staplesenergy}
\end{table}

\vspace{5cm}

\begin{table}[h!]
\scriptsize
    \centering
    \begin{tabular}{|l|l|l|l|}
        \hline
        \textbf{Ticker} & \textbf{Name} & \textbf{Sector} & \textbf{Sub-Industry} \\ \hline
ADP & Automatic Data Process & Industrials & HR \& Employment Serv \\  \hline
CSCO & Cisco & IT & Communications Equip \\  \hline
CSX & CSX Corporation & Industrials & Rail Transportation \\  \hline
CTAS & Cintas & Industrials & Diversified Support Serv \\  \hline
CTSH & Cognizant & IT & IT Consulting \\  \hline
FAST & Fastenal & Industrials & Trading Comp \& Distributors \\  \hline
HON & Honeywell & Industrials & Industrial Conglomerates \\  \hline
KDP & Keurig Dr Pepper & Cons Staples & Soft Drinks \& Beverages \\  \hline
LIN & Linde plc & Materials & Industrial Gases \\  \hline
MNST & Monster Beverage & Cons Staples & Soft Drinks \& Beverages \\  \hline
ORLY & O'Reilly Automotive & Cons Disc & Automotive Retail \\  \hline
PAYX & Paychex & Industrials & HR \& Employment Serv \\  \hline
PCAR & Paccar & Industrials & Const Mach \& Transp Equip \\  \hline
ROP & Roper Technologies & IT & Electronic Equip \& Instruments \\  \hline
SBUX & Starbucks & Cons Disc & Restaurants \\  \hline
TMUS & T-Mobile US & Comm Serv & Wireless Telecomm Serv \\  \hline
VRSK & Verisk Analytics & Industrials & Research \& Consulting Serv \\  \hline
\end{tabular}
    \caption{Level3 Cluster: Industrials}
    \label{tab:level3_industrials}
\end{table}

\newpage

\begin{table}[h!]
\centering
\scriptsize
\begin{tabular}{|l|l|c|c|c|c|}
\hline
\textbf{Level} & \textbf{Subgraph} & \textbf{N} & \textbf{Surgery links} & \textbf{G (iter)} & \textbf{Communities} \\ \hline
\multicolumn{6}{|l|}{\textit{Tech branch}} \\ \hline
Level 1 & Full NASDAQ-100 & 101 & 2310 & 25.03 (18) & 2 \\ \hline
Level 2 & Tech cluster & 66 & 800 & 69.87 (16) & 2 \\ \hline
Level 2b & Semiconductor sub-cluster & 16 & 60 & 19.48 (17) & 2 \\ \hline
Level 3 & 50-stock subgraph & 50 & 400 & 51.65 (20) & 2 \\ \hline
Level 4 & 40-stock subgraph & 40 & 204 & 33.70 (22) & 2 \\ \hline
Level 5 & 34-stock subgraph & 34 & 264 & 17.70 (18) & 2 \\ \hline
\multicolumn{6}{|l|}{\textit{Non-cyclical branch}} \\ \hline
Level 2 & Non-cyclical cluster & 35 & 250 & 44.94 (18) & 2 \\ \hline
Level 3 & 25-stock non-cyclical & 25 & 136 & 23.07 (20) & 2 \\ \hline
Level 4 & 17-stock industrials & 17 & 42 & 16.09 (11) & 2 \\ \hline
\end{tabular}
\caption{Curvature gap ratio $G = [\mu(\kappa^+) - \mu(\kappa^-)] / \sigma_{pooled}$ at each recursive hierarchical level of the NASDAQ-100 decomposition ($\alpha=0.5$, $G_{\min}=15$). The iteration count at which ARCUS stopped is shown in parentheses next to $G$. After Ricci flow, the curvature distribution separates into a high-curvature population $\kappa^+ \approx 0.5$ (intra-community links) and a low-curvature population $\kappa^- \approx 0$ (inter-community links). $\sigma_{pooled}$ is the pooled standard deviation of both populations. $G \gg 1$ at every level confirms that the bimodal separation is sharp and the surgery threshold self-selecting. This is an unsupervised analogue of the curvature gap of \cite{gosztolai2021unfolding} and \cite{fesser2024afrc}, which requires knowledge of the ground truth partition; here the post-flow curvature population replaces the known community labels.}
\label{tab:curvature_gap}
\end{table}

\newpage

\appendices
\section*{Appendix A: Wasserstein Distance and Probability Measure}

Consider a graph $G=(V,E,w)$ with vertices $V$, edges $E$ and weights $w$, with the degree of vertex $v$ indicated by $d_v$. We distribute a mass $m_x$ on each of the neighbors of the node $x$. Now we use the discrete Optimal Transport formulation to define a transport plan, the map $P: V \times V \rightarrow [0,1]$ such that $P(u,v)$ is the amount of mass at vertex $v$ to be moved to vertex $u$. 

\begin{align}
    \Sigma_{v^{\prime} \in V} P(u,v^{\prime}) = m_x(u) \\
 \Sigma_{u^{\prime} \in V} P(u^{\prime},v) = m_y(v)
\end{align}

the Wasserstein distance $W(m_x,m_y)$ is defined as the minimum total weighted distance to move $m_x$ to $m_y$
\begin{equation}
 W(m_x,m_y) = \min_{P}  \Sigma_{u,v \in V} d(u,v)P(u,v) 
\end{equation}

 The probability distribution $m_x$ for node $x$ must be specified. There are several possibilities to choose this distribution. Here, the probability distribution used for $x$, $m_x^{\alpha}$, with parameter $\alpha \in [0,1]$, with the set of neighbors of $x$, $\pi(x)$ is given by
 \begin{equation}
 m_x^{\alpha} (x_i) = 
\begin{cases}
 \alpha & \text{if $x_i = x$} \\
 \frac{1-\alpha}{d_{x}} & \text{if $x_i \in  \pi(x)$} \\
 0 & \text{otherwise}
\end{cases}
 \end{equation}
 Here $d_x$ is the normalization factor given by the degree of $x$ and $\alpha$ measures the probability to remain at $x$. In practice, it corresponds to distributing a unit mass over the site $x$ and all its nearest neighbors, with $\alpha$ on $x$ and $1-\alpha$ uniformly divided among the neighbors of $x$. This can be interpreted as a lazy random walk with $\alpha$ representing the probability of remaining at the site $x$. In the literature, $\alpha$ is known as the idleness parameter, with $\alpha=0$ representing the random walk and $\alpha=1$ the fully idle case. We define the $\alpha$-Ricci curvature, $\kappa^\alpha_{xy}$ as
 \begin{equation}
\kappa^\alpha_{xy} = 1-  \frac{W(m^\alpha_x,m^\alpha_y)}{d(x,y)} 
 \end{equation}

 For $\alpha=1$, we see that the curvature $\kappa_{xy}= 0$. Lin, Lu and Yau \citeyearpar{Yautohoku} introduced a new limiting Ollivier-Ricci curvature, $\bar\kappa$
 \begin{equation}
\bar\kappa = \lim\limits_{\alpha \rightarrow 1}  \frac{\kappa^\alpha_{ij}}{1-\alpha}
 \end{equation}
 They showed that the limit $\alpha\rightarrow 1$, of this quantity exists. In this paper, we use the value of $\alpha=1/2$, as widely used in the literature \cite{ni2019community} and refer to the $\alpha$-Ricci curvature as the Ricci curvature.

For our weighted complete graph we use a weight-proportional measure rather than the uniform measure above: $m_i(j) = (1-\alpha) \cdot w_{ij} / \sum_{k \neq i} w_{ik}$ for $j \neq i$, with $m_i(i) = \alpha$. This defines a lazy random walk in which the probability of hopping from $i$ to $j$ is proportional to the correlation strength between stocks $i$ and $j$ --- the natural financial interpretation. We have verified that the top-level split and main sub-clusters are identical under both the uniform and weight-proportional measures; differences appear only at the finest hierarchical levels among the smallest clusters.

\noindent\textbf{Weighted graph Laplacian.} For a weighted graph $G=(V,E,w)$ with edge weights $w_{ij}$, the weighted degree of node $i$ is $d_i = \sum_{j \neq i} w_{ij}$. The weighted graph Laplacian is
\begin{equation}
L = D - W
\end{equation}
where $D = \mathrm{diag}(d_1, \ldots, d_N)$ is the weighted degree matrix and $W$ is the weight matrix with entries $W_{ij} = w_{ij}$. The eigenvalues of $L$ are real and non-negative; the smallest is always 0 (with eigenvector $(1,\ldots,1)$, the market mode), and the second smallest $\lambda_2$ (the Fiedler value) controls the mixing time of the random walk and the connectivity of the graph. In the special case of equal edge weights $w_{ij} = c$ for all $i \neq j$ (the degenerate correlation matrix), $L$ reduces to $c(NI - J)$ where $J$ is the all-ones matrix, with eigenvalues 0 and $cN$ (multiplicity $N-1$) --- the degenerate ORC-spherical case discussed in Appendix~E.

\section*{Appendix B: ARCUS Algorithm Pseudocode}
\label{app:pseudocode}

We give a precise statement of the ARCUS (Adaptive Ricci Curvature Unsupervised Surgery) algorithm addressing the input graph, curvature computation, update rule, stopping rule, surgery rule, and recursion rule. The implementation uses the \texttt{GraphRicciCurvature} library \cite{github} with the OTD (Optimal Transport Distance) method.

\medskip
\noindent\textbf{Input.}
\begin{itemize}
  \item Return time series matrix $R \in \mathbb{R}^{T \times N}$ for $N$ stocks over $T$ trading days.
  \item Idleness parameter $\alpha = 1/2$.
  \item Maximum iterations $L_{\max} = 20$.
  \item Curvature gap threshold $\tau$ (determined empirically; see below).
  \item Minimum subgraph size $N_{\min} = 10$.
\end{itemize}

\medskip
\noindent\textbf{Step 1 --- Build weighted graph.}
Compute Pearson correlation matrix $C_{ij} = \mathrm{Corr}(R_i, R_j)$.
Set edge weight $w_{ij}^{(0)} = \sqrt{2(1 - C_{ij})}$ for all $i \neq j$.
Let $G^{(0)}$ be the complete weighted graph on $N$ nodes with these weights.

\medskip
\noindent\textbf{Step 2 --- Verify ORC-spherical condition.}
Compute $\kappa_{ij}^{(0)}$ for all edges using the weight-proportional measure (Appendix A).
Assert $\kappa_{ij}^{(0)} > 0$ for all $(i,j)$. If any $\kappa_{ij}^{(0)} \leq 0$, the graph is not ORC-spherical and the procedure does not apply.

\medskip
\noindent\textbf{Step 3 --- Ricci flow iteration.}
For $l = 0, 1, \ldots, L_{\max} - 1$:
\begin{enumerate}
  \item[(a)] Compute $\kappa_{ij}^{(l)}$ for all edges $(i,j)$.
  \item[(b)] Update: $\tilde{w}_{ij}^{(l+1)} = (1 - \kappa_{ij}^{(l)}) \cdot w_{ij}^{(l)}$.
  \item[(c)] Normalise: $w_{ij}^{(l+1)} = \tilde{w}_{ij}^{(l+1)} \cdot \bar{w}^{(0)} / \bar{w}^{(l+1)}$, where $\bar{w}^{(l)} = \frac{1}{|E|}\sum_{(i,j)} w_{ij}^{(l)}$ is the mean edge weight at step $l$. This normalisation preserves the mean edge weight at its initial value and prevents numerical collapse.
\end{enumerate}

\medskip
\noindent\textbf{Stopping rule.}
The iteration count is determined by an adaptive $G$-stability criterion: the flow runs until $G > G_{\min} = 15$ and the surgery set $n_{\mathrm{low}}$ has not changed for three consecutive iterations. This replaces a fixed $L_{\max}$, which is insufficient for deeper levels where the subgraph requires more iterations to converge. At Level~1 (N=101) the criterion fires at approximately 20 iterations; at Level~4 (N=40) approximately 36 iterations are required. As a visual check, we also inspect the curvature scatter plot (curvature vs.\ edge weight) at iterations 5, 10, 15 and 20 (Fig.~\ref{fig:Nasdaq100Level1RicciFlow}). Table~\ref{tab:G_convergence} reports $G$ at 5, 10 and 20 iterations for the tech branch. Theoretical support for running to convergence before a single surgery (rather than surgery at each iteration) is provided by Ma and Yang \citeyearpar{mayang2025}, who prove global existence and uniqueness for a modified flow $dw/dt = -\kappa\rho$ on arbitrary weighted graphs; their flow and curvature definition differ from ours, so the support is by analogy.

\medskip
\noindent\textbf{Step 4 --- Surgery.}
At the stopping iteration $L^*$ (Step 3), compute the final curvature $\kappa_{ij}^{(L^*)}$ for all edges. Under the normalised flow, the curvature distribution is bimodal: a low-$\kappa$ population (inter-community links, converging to $\kappa \approx 0$) and a high-$\kappa$ population (intra-community links, converging to $\kappa \approx \alpha \cdot N/(N-1)$), separated by an empty gap.

\noindent The surgery set is defined as the lower mode of this bimodal distribution:
\begin{equation}
S = \bigl\{(i,j) : \kappa_{ij}^{(L^*)} < \kappa^* \bigr\}
\end{equation}
where $L^*$ is the adaptive stopping iteration (the smallest $l$ such that $G^{(l)} > G_{\min}$ and the surgery set is stable across three consecutive iterations), and $\kappa^*$ is any value in the empty gap between the two modes. The curvature gap ratio
\begin{equation}
G(\kappa^*) = \frac{\mu(\kappa^+) - \mu(\kappa^-)}{\sigma_{pooled}}
\end{equation}
is constant across all $\kappa^*$ in the gap (Appendix C), confirming that $S$ is uniquely defined without any tuning: the surgery set is the lower population regardless of where in the gap $\kappa^*$ is placed. In the NASDAQ-100 data the gap spans approximately $\kappa \in [0.05, 0.45]$ at all hierarchical levels, and the surgery set contains the same edges for any $\kappa^*$ in this interval.

\noindent Remove all edges in $S$. Let $\mathcal{C}_1, \mathcal{C}_2, \ldots, \mathcal{C}_k$ be the connected components of $G \setminus S$.

\medskip
\noindent\textbf{Step 5 --- Recursion.}
For each component $\mathcal{C}_m$ with $|\mathcal{C}_m| \geq N_{\min}$:
\begin{enumerate}
  \item[(a)] Extract the submatrix $C[\mathcal{C}_m]$ of the original correlation matrix.
  \item[(b)] Construct the complete weighted subgraph $G_m^{(0)}$ on $|\mathcal{C}_m|$ nodes with weights $w_{ij} = \sqrt{2(1 - C_{ij})}$.
  \item[(c)] Verify that $G_m^{(0)}$ is ORC-spherical (Step 2). On the NASDAQ-100, each subgraph produced by surgery has been observed to satisfy this condition.
  \item[(d)] Apply Steps 3--5 to $G_m^{(0)}$.
\end{enumerate}
Recursion terminates when no component exceeds $N_{\min}$ or when surgery produces no disconnection (the flow converges to a uniformly positive curvature state with no bimodal separation, $G < G_{\min} = 2$).

\medskip
\noindent\textbf{Output.}
A hierarchical partition $\mathcal{P}_1 \supset \mathcal{P}_2 \supset \cdots$ of the $N$ stocks at successive recursive levels, together with the curvature gap ratio $G$ at each level (Table~\ref{tab:curvature_gap}).

\section*{Appendix C: Curvature Gap Ratio and Surgery Uniqueness}
\label{app:curvaturegap}

The curvature gap was introduced by Gosztolai and Arnaudon \citeyearpar{gosztolai2021unfolding} and used by Fesser et al.\ \citeyearpar{fesser2024afrc} as a supervised measure of how well a curvature separates known within-community from between-community edges. It is defined as:
\begin{equation}
\Delta = \frac{\mu(\kappa_{intra}) - \mu(\kappa_{inter})}{\sigma_{pooled}}
\end{equation}
where $\mu(\kappa_{intra})$ and $\mu(\kappa_{inter})$ are the mean curvatures of within- and between-community edges respectively, and $\sigma_{pooled} = \sqrt{[(\sigma_{intra})^2 + (\sigma_{inter})^2]/2}$ is the pooled standard deviation. Computing $\Delta$ requires knowledge of the ground truth partition.

Since our method operates on a ORC-spherical graph where communities are entirely latent prior to surgery, we introduce an unsupervised analogue. After $L_{\max}$ iterations of normalised Ricci flow on our NASDAQ-100 graph, the curvature distribution separates visually into two well-separated populations:
\begin{itemize}
  \item A \emph{high-curvature population} $\kappa^+$: intra-community links converging to $\kappa \approx 0.5 \approx \alpha \cdot N/(N-1)$, consistent with the ORC-spherical value of the initial complete graph.
  \item A \emph{low-curvature population} $\kappa^-$: inter-community links converging to $\kappa \approx 0$, the flat equilibrium of the normalised flow.
\end{itemize}
The two populations are separated by the threshold $\tau$ defined in Appendix B. We define the \emph{curvature gap ratio}:
\begin{equation}
G = \frac{\mu(\kappa^{+}) - \mu(\kappa^{-})}{\sigma_{pooled}}
\end{equation}
where $\mu(\kappa^+)$ is the mean curvature of edges with $\kappa_{ij} \geq \tau$, $\mu(\kappa^-)$ is the mean curvature of edges with $\kappa_{ij} < \tau$, and $\sigma_{pooled} = \sqrt{[(\sigma^+)^2 + (\sigma^-)^2]/2}$. In our data, $\mu(\kappa^+) \approx 0.5$ and $\mu(\kappa^-) \approx 0$, so $G \approx 0.5/\sigma_{pooled}$. Since both populations are tightly clustered after convergence, $\sigma_{pooled}$ is small and $G \gg 1$.

This is directly comparable to the Gosztolai-Arnaudon formula: the post-flow curvature population replaces the known community labels, making the measure unsupervised. $G \gg 1$ at every hierarchical level confirms that the bimodal separation is sharp and the surgery threshold $\tau$ is self-selecting --- the surgery set is insensitive to the precise value of $\tau$ within the natural gap.

We report $G$ at each recursive level in Table~\ref{tab:curvature_gap}.


The two populations are individually tight as well as well-separated. Across all five hierarchical levels, $\sigma(\kappa^+) \in [0.002, 0.021]$ and $\sigma(\kappa^-) \in [0.0004, 0.010]$, so the pooled standard deviation $\sigma_{pooled} \in [0.003, 0.015]$ is small relative to the gap $\mu(\kappa^+) - \mu(\kappa^-) \approx 0.47$. As a consequence, the curvature histogram is bimodal with an essentially empty interval between the two peaks --- any threshold $\tau \in [0.05, 0.45]$ identifies exactly the same surgery set at every level. The surgery criterion is therefore not merely robust to small perturbations in $\tau$: it is entirely insensitive to the choice of $\tau$ over a wide range.

\section*{Appendix D: Robustness Checks}
\label{app:robustness}

\noindent\textbf{Surgery threshold.}
The surgery set $S$ is the lower mode of the bimodal curvature distribution after Ricci flow. It requires no threshold parameter: the curvature gap ratio $G(\kappa^*)$ is constant for all $\kappa^*$ in the empty gap between the two modes (Appendix C). In the NASDAQ-100 data, $G$ takes the same value --- and $S$ contains the same edges --- for any $\kappa^* \in [0.05, 0.45]$ at every hierarchical level. The surgery criterion is therefore parameter-free.

\medskip
\noindent\textbf{Convergence of $G$ with iteration count.}
Table~\ref{tab:G_convergence} reports the curvature gap ratio $G$ at 10 iterations and at the ARCUS stopping iteration for each hierarchical level, computed on the NASDAQ-100 correlation matrix.

\begin{table}[h!]
\centering
\scriptsize
\begin{tabular}{|l|c|c|c|}
\hline
\textbf{Level} & \textbf{N} & \textbf{G (10 iter)} & \textbf{G at stop (iter)} \\ \hline
Level 1 (full) & 101 & 2.57 & 25.03 (18) \\ \hline
Level 2 (tech) & 66  & 3.06 & 69.87 (16) \\ \hline
Level 2b (semicond) & 16 & 4.83 & 19.48 (17) \\ \hline
Level 3 (50-stock) & 50  & 3.10 & 51.65 (20) \\ \hline
Level 4 (40-stock) & 40  & 4.30 & 33.70 (22) \\ \hline
Level 5 (34-stock) & 34  & 4.64 & 17.70 (18) \\ \hline
\end{tabular}
\caption{Curvature gap ratio $G$ at 10 iterations and at the ARCUS adaptive stopping point ($\alpha=0.5$, $G_{\min}=15$, stable window = 3 iterations). The iteration count at stopping is shown in parentheses. At 10 iterations $G$ is below $G_{\min}=15$ at every level, confirming the bimodal separation has not yet developed sufficiently for surgery. The adaptive stopping criterion fires between iterations 16 and 22 across levels, with Level 2 (tech) converging fastest and the 40-stock subgraph requiring the most iterations.}
\label{tab:G_convergence}
\end{table}

\noindent\textbf{Varying the idleness parameter $\alpha$.}
Since $\kappa^\alpha_{xy} \approx (1-\alpha)\,\bar{\kappa}_{xy}$ for $\alpha$ close to 1, varying $\alpha$ rescales all curvature values by $(1-\alpha)$ but preserves their ratios and therefore the bimodal gap structure and surgery set. This theoretical expectation predicts that the hierarchy should be stable for a wide range of $\alpha$, with variation appearing only where the approximation breaks down (small $\alpha$) or where curvature values become too small for the $G$-stability criterion to fire efficiently (large $\alpha$). We reran the full adaptive curvature-gap clustering with $\alpha \in \{0.25, 0.3, 0.4, 0.5, 0.75\}$. Table~\ref{tab:alpha_robustness} reports the surgery split sizes at each hierarchical level. The hierarchy is identical for $\alpha \in \{0.3, 0.4, 0.5, 0.75\}$ at the first three tech-branch levels. The healthcare/staples cluster and the Big Tech core (AAPL, ADBE, AMD, AMZN, ANSS, CDNS, Alphabet, META, MSFT, NVDA, SNPS) are identical for $\alpha \in \{0.3, 0.4, 0.5, 0.75\}$; at $\alpha = 0.25$ INTU joins Big Tech and the semiconductor cluster loses 7 stocks. Values for $\alpha \geq 0.9$ are excluded: near the Lin-Lu-Yau limit $\alpha \to 1$, curvature values scale as $(1-\alpha)\bar{\kappa} \to 0$, requiring prohibitively many iterations for the $G$-stability criterion to fire. These results confirm that the main economic groupings are robust over a wide range of $\alpha$; variation appears only at the deepest recursive levels, consistent with the theoretical expectation that $\alpha$ rescales curvature values but preserves the bimodal gap structure.

\begin{table}[h!]
\centering
\scriptsize
\begin{tabular}{|l|c|c|c|c|c|}
\hline
\textbf{Level} & \textbf{$\alpha=0.25$} & \textbf{$\alpha=0.3$} & \textbf{$\alpha=0.4$} & \textbf{$\alpha=0.5$} & \textbf{$\alpha=0.75$} \\ \hline
L1: tech / non-cyclical & 67 / 34 & 66 / 35 & 66 / 35 & 66 / 35 & 66 / 35 \\ \hline
L2: semiconductors / rest & 16 / 51 & 16 / 50 & 16 / 50 & 16 / 50 & 16 / 50 \\ \hline
L3: cross-sector / rest & 9 / 42 & 9 / 41 & 10 / 40 & 10 / 40 & 10 / 40 \\ \hline
L4: biotech / rest & 5 / 37 & 4 / 37 & 6 / 34 & 6 / 34 & 6 / 34 \\ \hline
L5: Big Tech / growth & 13 / 24 & 14 / 23 & 12 / 22 & 12 / 22 & 12 / 10$^\dagger$ \\ \hline
Non-cyc L2: health-staples / rest & 10 / 24 & 10 / 25 & 10 / 25 & 10 / 25 & 9 / 16 \\ \hline
Non-cyc L3: staples-energy / industrials & 8 / 16 & 8 / 17 & 8 / 17 & 8 / 17 & --- \\ \hline
\end{tabular}
\caption{Surgery split sizes (larger / smaller component) at each hierarchical level for five values of $\alpha$. The healthcare/staples cluster (10 stocks) and Big Tech core (12 stocks) are identical for $\alpha \in \{0.3, 0.4, 0.5, 0.75\}$. At $\alpha=0.25$, INTU joins Big Tech and the semiconductor cluster contracts by 7 stocks. $^\dagger$At $\alpha=0.75$ the 22-stock growth cluster splits further into 12+10 at this level. Values for $\alpha \geq 0.9$ are excluded: near the Lin-Lu-Yau limit $\alpha \to 1$, curvature values scale as $(1-\alpha)\bar{\kappa} \to 0$, requiring prohibitively many iterations to reach the $G$-stability stopping criterion.}
\label{tab:alpha_robustness}
\end{table}

\noindent\textbf{Rolling window analysis.}
A rolling window analysis of the NASDAQ-100 correlation structure would inevitably reflect changes in the underlying economy — the COVID-19 crash, the post-pandemic recovery, and sector rotation across regimes — rather than testing the stability of the ARCUS algorithm itself. Disentangling methodological stability from genuine economic regime changes requires a more detailed econophysical analysis that is the subject of a companion paper \cite{Bhargavicluster}. We note that the $\alpha$-robustness results above provide a more direct stability check: they show the hierarchy is insensitive to the choice of random walk measure over a wide range of $\alpha$ values, which perturbs the underlying geometry of the curvature computation rather than merely the time period of the data.

\section*{Appendix E: Geometric Foundations}
\label{app:geometry}

\noindent\textbf{E.0 The equal-weight complete graph.}
For a complete graph $K_n$ with $N$ nodes and equal edge weights, with $\alpha=1/2$, all Ollivier-Ricci curvatures are equal. The Wasserstein distance calculation (illustrated in Fig.~\ref{fig:wasssphere} for a 5-node graph) gives $W_{ij} \leq 1$, so the Ricci curvature is always positive:
\begin{equation}
\kappa_{ij} = (1-\alpha)\frac{n}{n-1}
\end{equation}
It is known \cite{Yautohoku} that the Lin-Lu-Yau curvature $\lim_{\alpha\to 1}\kappa_{ij}/(1-\alpha) = n/(n-1)$ of $K_n$ is the unique graph with constant curvature greater than 1. The eigenvalues of the unweighted graph Laplacian of $K_n$ are $0$ and $n/(n-1)$ with multiplicity $n-1$, coinciding with the Lin-Lu-Yau curvatures. For $n=101$ and $\alpha=1/2$, this gives $\kappa = 0.505$: the maximally degenerate ORC-spherical case.

Under Ricci flow, this graph evolves self-similarly: all edges shorten at equal rates, curvature remains uniform, and the graph collapses to a point. In the normalised variant used by GraphRicciCurvature \cite{github}, mean edge weight is preserved and collapse is prevented, but self-similarity remains throughout --- the graph stays complete with equal weights and no community separation is possible. The neckpinch that develops in the empirical NASDAQ-100 graph is driven entirely by the heterogeneity of the empirical weights: the degeneracy of the equal-weight baseline is broken by the spread of the correlation matrix. This connection to the market mode of Laloux et al.\ \citeyearpar{Laloux_1999} --- in which a single common factor drives all correlations equally --- makes clear why the equal-weight graph is the natural baseline and why departures from it are geometrically significant.

\noindent\textbf{Hamilton's theorem and the ORC-spherical graph.}
Hamilton \citeyearpar{hamilton1982three} proved that on any compact, simply-connected three-manifold with strictly positive Ricci curvature, the Ricci flow converges (after rescaling) to a round sphere. This is the foundational result of Ricci flow theory. Our NASDAQ-100 correlation network is motivated by this ORC-spherical setting: all initial Ollivier-Ricci curvatures are strictly positive (Definition 1 in the main text).

The ORC-spherical case is, in a precise sense, the fundamental case for Ricci flow community detection. Prior network community detection work began from mixed-curvature graphs --- SBM synthetic networks \cite{ni2019community} and MST-filtered financial networks \cite{samal2021} --- where community structure was pre-encoded in the initial geometry and the Ricci flow amplified a pre-existing signal. Starting from an all-positive ORC-spherical graph requires the flow to generate structure from scratch, more faithfully mirroring Hamilton's original setting. Financial correlation networks naturally produce all-positive initial curvatures, making this the practically relevant case.

\noindent\textbf{The neckpinch singularity.}
The term `neckpinch singularity' originates in differential geometry. Hamilton \citeyearpar{hamilton1995singularities} identified it as one of the possible singularity types under Ricci flow. Angenent and Knopf \citeyearpar{angenent2004} gave the first rigorous construction on a rotationally symmetric dumbbell on $S^{n+1}$, showing the neck region has locally cylindrical geometry $S^{n-1} \times \mathbb{R}$. The cylinder has zero Ricci curvature in the $\mathbb{R}$ direction while remaining positively curved in the $S^{n-1}$ cross-sections --- it is the natural intermediate geometry between the positively curved ($\kappa > 0$) spherical ends and the hyperbolic ($\kappa < 0$) regime that does not appear to develop in the ORC-spherical case. Surgery at the neckpinch therefore cuts along zero-curvature edges; in the discrete setting this is precisely the lower population of the bimodal curvature distribution.

Ni et al.\ \citeyearpar{ni2019community} imported the term `neckpinch singularity' into the network context by analogy. In their sparse unweighted networks (Karate Club, Facebook ego network with 792 nodes and 24 hand-labelled friend circles, SBM synthetic networks), inter-community edges already carry negative Ollivier-Ricci curvature before any flow is applied --- the neckpinch is identified from the pre-existing negative curvature structure rather than from first principles. Our work provides this identification from a ORC-spherical initial state: after sufficient iterations of Ricci flow, the curvature distribution polarises into a positively curved population ($\kappa > 0$, intra-community links) and a near-zero population ($\kappa \approx 0$, cylindrical, surgery links). The two populations are sharply separated by an empty gap, with no hyperbolic ($\kappa < 0$) edges --- consistent with the ORC-spherical initial condition being preserved in the intra-community subgraphs.

\noindent\textbf{The discrete neckpinch and Hamilton's theorem: an open question.}
After Ricci flow on our ORC-spherical graph, two geometric regimes emerge, corresponding to the spherical and cylindrical geometries:
\begin{itemize}
\item \emph{Positive curvature} ($\kappa > 0$): spherical edges --- intra-community links, high neighbourhood overlap, shrink under the flow.
\item \emph{Near-zero curvature} ($\kappa \approx 0$): cylindrical edges --- the discrete neckpinch, surgery links. These are the inter-community edges that approach the curvature signature of cylindrical geometry $S^{n-1} \times \mathbb{R}$ as the dumbbell neck develops.
\end{itemize}
We emphasise that $\kappa \approx 0$ here identifies \emph{cylindrical} geometry, not flat geometry. The regular grid is flat and has $\kappa = 0$, but the cylinder $S^{n-1} \times \mathbb{R}$ also has zero curvature along its axis --- one principal curvature vanishes while the cross-sectional sphere remains positively curved. The surgery links at $\kappa \approx 0$ are the discrete analogue of the cylindrical neck, not of a flat grid. The bimodal curvature distribution (intra-community at $\kappa \approx 0.48$, surgery at $\kappa \approx 0$) is the discrete analogue of the dumbbell: two spherical ends connected by a cylindrical neck.

The absence of hyperbolic ($\kappa < 0$) edges after the flow is consistent with the ORC-spherical initial condition; this contrasts with sparse unweighted graphs where inter-community edges already carry negative curvature before any flow is applied (\cite{ni2019community}). With $\alpha = 1/2$, the initial curvatures satisfy $\kappa > \alpha = 0.5$ for a complete graph; after flow the surgery population converges toward $\kappa \approx 0$, well below the initial baseline, while the intra-community population remains near $\kappa \approx 0.48$.

\emph{Open question. Whether the surgery links approach $\kappa = 0$ from above without crossing into the hyperbolic regime is discussed in the main text. Resolving it theoretically would require either higher-precision optimal transport solvers or a monotonicity result for the discrete Ricci flow on weighted ORC-spherical graphs --- a discrete analogue of Hamilton's theorem.}

\noindent\textbf{Thurston's geometrisation conjecture and recursive surgery.}
Thurston \citeyearpar{thurston2} conjectured that every compact 3-manifold decomposes into pieces each admitting one of eight canonical geometric structures. Perelman proved this conjecture using Ricci flow with surgery: neckpinch surgery decomposes the manifold into simpler pieces, each converging under continued flow to its canonical Thurston geometry. Our recursive community detection procedure is the discrete financial analogue: each round of surgery decomposes the correlation network into spherical communities, each undergoing its own Ricci flow from an initial state observed to be ORC-spherical on the NASDAQ-100. The hierarchical levels of our NASDAQ-100 decomposition correspond to successive geometric surgeries.

\noindent\textbf{The Lin-Lu-Yau programme and Hamilton's theorem.}
Lin, Lu and Yau \citeyearpar{Yautohoku} developed discrete analogues of Hamilton's positive curvature theorems --- Bonnet-Myers, Lichnerowicz, and Cheng --- providing the theoretical scaffolding for Ricci flow on graphs with positive initial curvature. The present paper can be read as following the spirit of Hamilton's programme in this discrete weighted setting: beginning from the ORC-spherical initial condition, the flow develops a neckpinch, surgery decomposes the graph into positively curved pieces, and the procedure recurses. The open question of whether inter-community links remain non-negative throughout (approaching $\kappa = 0$ from above rather than becoming hyperbolic) is the discrete weighted analogue of Hamilton's theorem --- if confirmed, it would show that Ricci flow on a positively curved weighted graph preserves positive curvature in the communities while developing a cylindrical neckpinch at $\kappa \approx 0$, without producing hyperbolic geometry.

The Ollivier-Ricci curvature is bounded by $-2$ and $1$ (\cite{Yaueigenvalues}, \cite{jostandliu}). The Lin-Lu-Yau lower bound $\bar{\kappa}_{ij} \geq 2/d_i + 2/d_j - 2$ (\cite{Yautohoku}) was proved for unweighted graphs with uniform measure ($\alpha=0$). For a complete unweighted graph on $n$ nodes this gives $\bar{\kappa}_{min} = 2(3-n)/(n-1)$; for $n=101$ this is $\bar{\kappa}_{min} \approx -1.96$. The $\alpha$-scaling $\kappa^\alpha \approx (1-\alpha)\bar{\kappa}$ (exact as $\alpha \to 1$, approximate at $\alpha=0.5$) gives $\kappa^{0.5}_{min} \approx -0.98$, corresponding to a maximum edge growth rate of $(1-\kappa^{0.5}) \approx 1.98$ per iteration. The observed surgery link weights of $\approx 3$ after 20 iterations are consistent with this estimate from an average initial weight of $1.11$. Both the LLY bound (unweighted) and the $\alpha$-scaling (valid near $\alpha \to 1$) are approximations in our setting; the precise extension to weighted graphs with $\alpha = 0.5$ remains open.

\section*{Appendix F: Mantegna Benchmark Validation}
\label{app:mantegna}

This appendix validates the Ricci flow community detection procedure against the ten-stock correlation network benchmark of Mantegna \citeyearpar{mantegna1999hierarchical} and Tumminello, Lillo and Mantegna \cite{tumminello2008}.

\noindent\textbf{F.1 Benchmark dataset.}
The benchmark consists of ten U.S.\ equities: AIG, IBM, BAC, AXP, MER (financial sector); TXN, MOT (technology sector); and SLB, OXY, RD (energy sector). The correlation matrix is taken directly from the original study. Using average-linkage cluster analysis on correlation-based distances, Tumminello et al.\ \citeyearpar{tumminello2008} recover a hierarchical tree in which these three sector groupings emerge as the dominant branches. All ten initial ORC values are positive (min $\kappa = 0.450$), confirming the graph is positively curved.

\noindent\textbf{F.2 Ricci flow result.}
Surgery after 20 iterations splits the complete 10-stock graph into two components: \{SLB, OXY, RD\} (energy) and \{AIG, IBM, BAC, AXP, MER, TXN, MOT\} (remaining 7 stocks). The energy sector splits off exactly; ARI = 1.000 against the two-community ground truth; $G = 128.9$.

Running Ricci flow on the 7-stock subgraph for 10 iterations and applying surgery recovers two communities: the financial cluster \{AIG, IBM, BAC, AXP, MER\} and the technology cluster \{TXN, MOT\}. $G = 25.9$ at this level. The full three-sector partition achieves ARI = 1.000, NMI = 1.000, $Q = 0.018$ against the Mantegna ground truth.

\noindent\emph{Note on iteration count.} Smaller subgraphs reach bimodal separation at fewer iterations. At 10 iterations the TXN-MOT intra-community edge remains at high curvature ($\kappa \approx 0.67$), clearly separated from the cross-community edges ($\kappa \approx 0.05$), giving $G = 25.9$. At 20 iterations the TXN-MOT edge over-converges to $\kappa = 0$, collapsing into the surgery population and producing singletons. This indicates that the optimal iteration count scales with subgraph size; a stopping criterion based on $G$ stability rather than a fixed $L_{\max}$ would be more principled for small subgraphs.

\noindent\textbf{F.3 Comparison with MST and PCA.}
We compare Ricci flow against MST and two natural PCA approaches, all parameter-free except where noted. The PCA sign method assigns communities by the sign of eigenvector loadings, directly analogous to the NVIDIA comparison in the main text: stocks with the same sign of $v_k$ belong to the same group, with no $k$-means and no threshold.

\begin{table}[h!]
\centering
\scriptsize
\begin{tabular}{|l|c|c|l|}
\hline
\textbf{Method} & \textbf{$k$} & \textbf{ARI} & \textbf{Note} \\ \hline
Ricci flow Levels 1+2 (this paper) & 3 & 1.000 & $G=128.9$, $25.9$; no parameters \\ \hline
MST / avg-linkage (Mantegna) & 3 & 1.000 & Requires specifying $k=3$ \\ \hline
PCA $v_2$ sign (parameter-free) & 2 & 0.566 & Energy splits correctly; technology/financial indistinguishable \\ \hline
PCA $v_2$+$v_3$ sign (parameter-free) & 3 & 0.566 & $v_3$ does not align with financial/technology boundary \\ \hline
\end{tabular}
\caption{ARI against the Mantegna three-sector ground truth \cite{tumminello2008}. The PCA sign method assigns communities by the sign of eigenvector $v_2$ (the leading non-market eigenvector, $\lambda_2 = 1.396$): positive loadings identify energy stocks (SLB, OXY, RD), negative loadings the remaining seven stocks. This gives a correct two-way split but cannot recover the technology/financial boundary, because $v_3$ ($\lambda_3 = 0.959$) does not align with this boundary --- IBM and TXN both load negatively on $v_3$, indistinguishable from the financial cluster. Ricci flow recovers both splits correctly at Level 1 (energy) and Level 2 (technology vs financial) without specifying the number of communities. MST recovers all three sectors but requires $k$ to be specified.}
\label{tab:mantegna_ari}
\end{table}

\begin{figure}[h!]
\begin{center}
\begin{minipage}{150mm}
\includegraphics[width=\columnwidth]{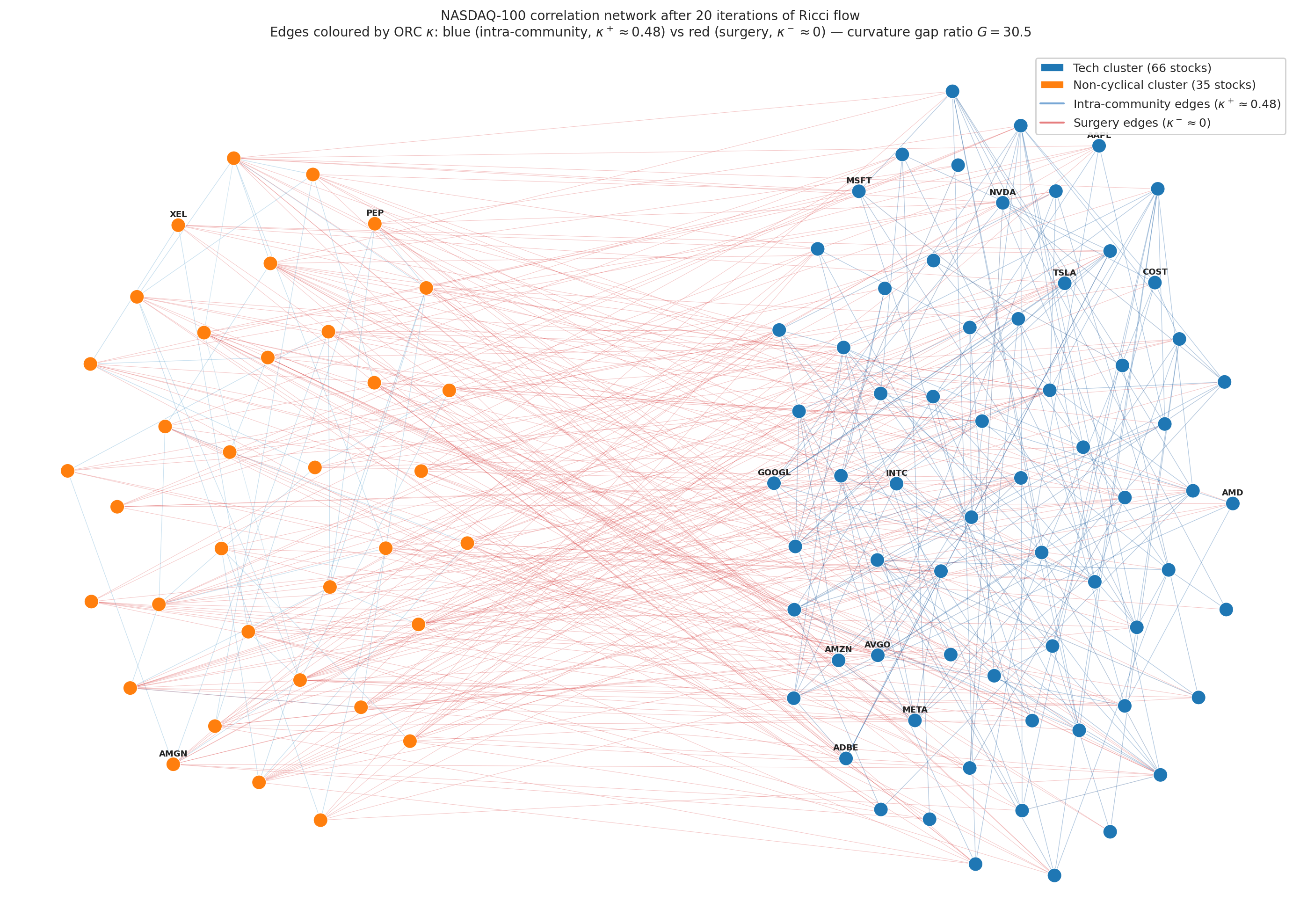}
\caption{NASDAQ-100 correlation network after 20 iterations of ARCUS ($\alpha=0.5$, Level 1). Nodes coloured by community: blue = tech cluster (66 stocks), orange = non-cyclical cluster (35 stocks). Edges coloured by ORC $\kappa$: blue edges are intra-community links ($\kappa^+ \approx 0.48$), red edges are surgery links ($\kappa^- \approx 0$). A subset of edges is shown for readability. The dumbbell geometry with a red neck of zero-curvature surgery links is the discrete analogue of the neckpinch singularity in Ricci flow on manifolds. Key stocks labelled.}
\label{fig:nasdaq_network}
\end{minipage}
\end{center}
\end{figure}

\end{document}